\documentclass{article}
\usepackage{amsmath}
\usepackage{amssymb}
\usepackage{amsfonts}
\usepackage{bm}
\usepackage[dvips,xdvi]{graphicx}
\usepackage{cite}

\newcommand{\erfc}{\operatorname{erfc}}

\begin{document}



\title{\bf Detailed Balance Condition and Effective Free Energy
in the Primitive Chain Network Model}
\author{Takashi Uneyama and Yuichi Masubuchi\\
\\
JST-CREST, Institute for Chemical Research, Kyoto University, \\
Gokasho, Uji, Kyoto 611-0011, Japan}

\maketitle



\begin{abstract}
We consider statistical mechanical properties of the
primitive chain network (PCN) model for entangled polymers from its
dynamic equations.
We show that the dynamic equation for the segment number of the PCN
model does not reduce to the standard Langevin equation which satisfies the
detailed balance condition.
We propose heuristic modifications for the PCN dynamic equation for the
segment number, to make it reduce to the standard Langevin equation.
We analyse some
equilibrium statistical properties of the modified PCN model, by using
the effective free energy obtained from the modified PCN dynamic equations.
The PCN effective free energy can be interpreted as the sum of the
ideal Gaussian chain free energy and the repulsive
interaction energy between slip-links. By using the single chain approximation,
we calculate several distribution functions of the PCN model.
The obtained distribution functions are
qualitatively different from ones for the simple
 slip-link model without any direct interactions between slip-links.
\end{abstract}


\section{Introduction}

The primitive chain network (PCN) model
\cite{Masubuchi-Takimoto-Koyama-Ianniruberto-Greco-Marrucci-2001,Masubuchi-Ianniruberto-Greco-Marrucci-2008}
is a slip-link type mesoscopic coarse-grained model for entangled
polymeric systems\cite{Doi-Edwards-book}.
To simulate rheological properties efficiently, the PCN model represents
entangled polymers as a network like structure of which topology
dynamically changes.
Simulations based on the PCN model can reproduce
rheological properties of entangled polymers well with relatively small
computational costs, and it has been applied various systems
including branched polymers\cite{Masubuchi-Ianniruberto-Greco-Marrucci-2006a,Masubuchi-Yaoita-Matsumiya-Watanabe-2011}
or bidisperse polymers\cite{Masubuchi-Watanabe-Ianniruberto-Greco-Marrucci-2008}.
Although the PCN model achieved success to predict various rheological
behaviours of entangled polymers, its statistical properties are still
not fully understood.

In recent years, various primitive path analysis methods\cite{Everaers-Sukumaran-Grest-Svaneborg-Sivasubramanian-Kremer-2004,Kroger-2005,Tzoumanekas-Theodorou-2006}
have been developed to extract statistical properties of
network structures in entangled polymers quantitatively from atomistic or coarse-grained
molecular models (such as the molecular dynamics simulations).
It is an interesting question whether the extracted statistical data
agree with ones obtained by the PCN model, to discuss consistency or
relation between the models.
Quite recently, the network statistics of the PCN model were systematically
examined\cite{Masubuchi-Uneyama-Watanabe-Ianniruberto-Greco-Marrucci-2010}.
The statistical properties of the PCN model are shown to be qualitatively in
agreement with primitive path analysis data.
However, the analysis is conducted only for simulation results,
and from the theoretical view point, how the network statistics
is determined in the PCN model is still not clear.
The PCN model is a dynamical model which is constructed in a rather
phenomenological way. This makes theoretical analysis for the equilibrium
statistics of the PCN model difficult.
Moreover, the thermodynamic validity of the PCN model is not guaranteed
from the view point of the statistical mechanics.

To achieve the thermal equilibrium state, the detailed balance condition is
required to be satisfied. If the detailed balance condition is satisfied
for a model, we can utilize the standard statistical mechanics to
analyze the model. Namely, the equilibrium probability distribution is
given by the Boltzmann type equilibrium distribution with the thermodynamic
potential. In the vicinity of the equilibrium state,
dynamical behaviours such as the relaxation or linear response
properties can be related to the correlation functions of fluctuations
in equilibrium\cite{Kubo-Toada-Hashitsume-book,Evans-Morris-book,Risken-book}.
The detailed balance
condition is a strong condition and thus it is not always satisfied in a
phenomenologically constructed dynamical model.
Therefore, it is desired to examine whether the PCN model satisfies the
detailed balance condition or not. So far, most of the previous works
for the PCN model focused on dynamical properties such as rheological properties.
As far as the authors know, the analysis of
the PCN model from 
the view point of statistical mechanics has never been shown explicitly 
in the literatures.
(It would be fair to mention that several researchers have noticed that the PCN
model or some other slip-link models do not satisfy the detailed balance
condition, although it has not been stressed in published literatures.
For some slip-link models, the statistical
mechanical analysis or the modeling consistent with the detailed balance
condition have been done\cite{Nair-Schieber-2006,Uneyama-2011}.)

In this work, first we attempt to interpret the dynamic equations of the
PCN model as the Langevin equations.
We show that the PCN dynamic equations do not satisfy the detailed
balance condition and therefore the PCN model does not have the thermal
equilibrium state. Then we propose possible modifications to recover the
detailed balance condition, in a heuristic way. Even if we
modify the PCN model, the resulting thermodynamic potential (effective
free energy) of the PCN model is not identical to the free energy of the
slip-link model without any direct interactions. In the PCN model, there is the
effective repulsive interaction 
between neighboring slip-links on a polymer chain. Based on the obtained
effective free energy and the single chain approximation, finally we
calculate several equilibrium probability distribution functions
analytically. We compare analytical results with other theories as well as
PCN simulation data.

\section{Statistical Mechanical Interpretation of PCN Dynamic Equations}
\label{dynamic_equations_for_the_primitive_chain_model}

\subsection{PCN Dynamic Equations}
\label{pcn_dynamic_equations}

In the PCN
model\cite{Masubuchi-Takimoto-Koyama-Ianniruberto-Greco-Marrucci-2001,Masubuchi-Ianniruberto-Greco-Marrucci-2008},
entangled polymers are represented as a network
structure which consists of nodes (slip-linked points and chain ends)
and bonds which connect nodes.
For simplicity we limit ourselves to monodisperse linear polymer systems
with the polymerization index (number of segments) being $N$.
The state of the system is described by the set of
coarse-grained variables; positions of
slip-linked nodes and end nodes, numbers of segments between two neighboring
nodes, and the connectivity information.
For convenience, we express the index of the $k$-th node in the $i$-th
chain as $(i,k)$.
We express the position of the $(i,k)$ node as $\bm{R}_{i,k}$, and
the number of segments between the $(i,k - 1)$ and $(i,k)$ nodes as
$N_{i,k}$.
We also express the number of subchains in the $i$-th chain as $Z_{i}$.
The $0$th and $Z_{i}$-th nodes in the $i$-th chain represent chain ends
whereas other nodes represent slip-linked nodes.
A slip-linked node is assumed to be spatially coupled to the partner
node. We express this by introducing a connectivity map for the node
index, $C$ (the $(i,k)$ node is coupled to the $C(i,k)$ node).
Two nodes connected by slip-links, $(i,k)$ and $C(i,k)$, share the same
node position.
\begin{equation}
 \label{connectivity_information_node_position}
 \bm{R}_{i,k} = \bm{R}_{C(i,k)} \qquad (\text{for all slip-linked nodes})
\end{equation}
The state of the system can be completely described by $\lbrace
\bm{R}_{i,k} \rbrace$, $\lbrace N_{i,k} \rbrace$, $\lbrace Z_{i}
\rbrace$, and $C$. (The information of $C$ is not important in the
following arguments and thus we do not describe it explicitly.)

In absence of external flow (deformation) field,
the PCN dynamic equations
\cite{Masubuchi-Ianniruberto-Greco-Marrucci-2008}
are described as follows. (Although there are several different versions
of the PCN model, in this work we employ the version described in Ref
\citen{Masubuchi-Ianniruberto-Greco-Marrucci-2008}.)
\begin{equation}
 \label{dynamic_equation_node_position}
 \zeta \frac{d\bm{R}_{i,k}(t)}{dt}
  = \frac{3 k_{B} T}{b^{2}} 
  \sideset{}{'} \sum_{j,l} \frac{\bm{R}_{j,l}
  - \bm{R}_{i,k}}{N_{j,l}}
  + \sqrt{2 k_{B} T \zeta } \bm{w}^{(\bm{R})}_{i,k}(t)
\end{equation}
\begin{equation}
 \label{dynamic_equation_segment_number}
  \frac{dN_{i,k}(t)}{dt}
  = 
  \begin{cases}
   - J_{i,1}(t) & (k = 1) \\
   - J_{i,k}(t) + J_{i,k - 1}(t) & (2 \le k \le Z_{i} - 1) \\
   J_{i,Z_{i} - 1}(t) & (k = Z_{i})
  \end{cases}
\end{equation}
\begin{equation}
 \label{dynamic_equation_segment_number_flux}
   \frac{\zeta}{2 \rho_{i,k}}
   J_{i,k}(t)
   \equiv
    \displaystyle \frac{3 k_{B} T}{b^{2}}
   \left[ \frac{|\bm{R}_{i,k + 1} - \bm{R}_{i,k}|}{N_{i,k + 1}}
   - \frac{|\bm{R}_{i,k} - \bm{R}_{i,k - 1}|}{N_{i,k}} \right]
   + \sqrt{k_{B} T \zeta} w^{(N)}_{i,k}(t)
\end{equation}
where $\zeta$ is the friction coefficient of a node,
$k_{B}$ is the Boltzmann constant, $T$ is temperature, $b$ is the
segment size,
and $J_{i,k}(t)$ is the flux of the segment number on the $(i,k)$ node. The summation in the
right hand side of eq \eqref{dynamic_equation_node_position} is taken
for all nodes connected to the target (topological neighbor nodes). The
topological neighbor nodes are $(i,
k \pm 1)$ and $(i',k' \pm 1)$ (with $(i',k')
= C(i,k)$) for a slip-linked node, and $(i, k + 1)$ or $(i, k - 1)$ for
an end node.
$\rho_{i,k}$ is the segment density along 
the polymer chain on the $(i,k)$ node, and is defined as the arithmetic average of local densities in
two neighboring bonds.
\begin{equation}
 \label{line_density_definition}
  \rho_{i,k}(\lbrace \bm{R}_{i,k} \rbrace, \lbrace N_{i,k} \rbrace) \equiv \frac{1}{2} \left[ \frac{N_{i,k}}{|\bm{R}_{i,k} - \bm{R}_{i,k-1}|} +
   \frac{N_{i,k + 1}}{|\bm{R}_{i,k + 1} - \bm{R}_{i,k}|}
   \right]
\end{equation}
$\bm{w}^{(\bm{R})}_{i,k}(t)$ and $w_{i,k}^{(N)}(t)$ are Gaussian white noises which
satisfy the following relations.
\begin{align}
 & \label{w_r_fdr} \langle \bm{w}^{(\bm{R})}_{i,k}(t) \rangle = 0, 
 \qquad \langle \bm{w}^{(\bm{R})}_{i,k}(t) \bm{w}^{(\bm{R})}_{j,l}(t') \rangle = \delta_{ij}
 \delta_{kl} \delta(t - t') \bm{1} \\
 & \label{w_n_fdr} \langle {w}^{(N)}_{i,k}(t) \rangle = 0, 
 \qquad \langle w^{(N)}_{i,k}(t) w^{(N)}_{j,l}(t') \rangle = \delta_{ij}
 \delta_{kl} \delta(t - t') \\
 & \label{w_r_n_fdr} \langle \bm{w}^{(\bm{R})}_{i,k}(t) w^{(N)}_{j,l}(t') \rangle = 0
\end{align}
where $\langle \dots \rangle$ means the statistical average and $\bm{1}$
is the unit tensor.

We note that the (osmotic) repulsive force terms
\cite{Masubuchi-Takimoto-Koyama-Ianniruberto-Greco-Marrucci-2001} are
dropped in the dynamic equations
\eqref{dynamic_equation_node_position}-\eqref{dynamic_equation_segment_number_flux}.
The repulsive force is required only to cancel the
artificial attractive interaction in multi chain slip-link systems
\cite{Uneyama-Horio-2011}, and its contribution is not essential in the analysis in this
work. (Actually, it is empirically known that
as long as the repulsion is sufficiently strong to avoid
aggregation, the PCN model shows almost the same statistics.)

To specify the PCN dynamics completely, we also need the network
reconstruction rules. Whether the network is reconstructed or not is
determined by the number of segments in chain end bonds.
If the number of segments in a chain end bond
becomes larger than a certain criterion, a new pair of slip-linked nodes
are constructed. On the other hand, if the number of the segments in an
chain end bond becomes smaller than another criterion, the slip-link
attached to that bond is destructed. The criteria are given as $3
N_{0} / 2$ and $N_{0} / 2$, with $N_{0}$ being the average number of
segments in a bond.
Although the network reconstruction rules affect
the statistics rather strongly\cite{Masubuchi-Uneyama-Watanabe-Ianniruberto-Greco-Marrucci-2010}, how they affect the statistics is not
so clear. (It is reported that other reconstruction rules
\cite{Yaoita-Isaki-Masubuchi-Watanabe-Ianniruberto-Greco-Marrucci-2008,Masubuchi-Uneyama-Watanabe-Ianniruberto-Greco-Marrucci-2010}
can improve several statistical properties.)
In this work we do not consider about the reconstruction process and
concentrate on the
dynamic equations for $\lbrace \bm{R}_{i,k} \rbrace$ and $\lbrace
N_{i,k} \rbrace$ (eqs \eqref{dynamic_equation_node_position}-\eqref{dynamic_equation_segment_number_flux}).

\subsection{Detailed Balance Condition in the PCN Model}
\label{langevin_form_of_segment_transport_equation_and_effective_free_energy}

Because the PCN model is designed as a dynamical model, its static
properties are not clear from its dynamic equations
\eqref{dynamic_equation_node_position}-\eqref{dynamic_equation_segment_number_flux}.
Moreover, the existence of the thermodynamic
potential (the free energy) is generally not guaranteed for such a
phenomenological dynamical model. If eqs
\eqref{dynamic_equation_node_position}-\eqref{dynamic_equation_segment_number_flux}
satisfy the detailed balance condition, 
the forces are expressed as variations of the free energy. (In the
followings, we call such forms as the variational forms.)
Then we can construct the free energy from eqs
\eqref{dynamic_equation_node_position}-\eqref{dynamic_equation_segment_number_flux}.
In this section we attempt to interpret the PCN dynamic equations as
Langevin equations\cite{Gardiner-book} which satisfy the detailed balance
condition.

According to the standard nonequilibrium statistical physics \cite{Kawasaki-1973,Sekimoto-book,Risken-book},
dynamics of the system in the vicinity of equilibrium
can be described well by the Langevin equations.
If we assume that there is no dynamic coupling between  $\lbrace
\bm{R}_{i,k} \rbrace$ and $\lbrace
N_{i,k} \rbrace$, nor the memory effect, the detailed-balanced Langevin
equations can be formally described as follows.
\begin{align}
 \label{langevin_equation_r_generic}
 & \frac{d\bm{R}_{i,k}(t)}{dt} = - \sum_{j,l} \bm{L}^{(\bm{R})}_{i,k;j,l}
 \cdot \frac{\partial \mathcal{F}_{\text{eff}}(\lbrace \bm{R}_{i,k} \rbrace,\lbrace N_{i,k}
  \rbrace, \lbrace Z_{i} \rbrace)}{\partial \bm{R}_{i,k}}
 + k_{B} T \sum_{j,l} \frac{\partial}{\partial \bm{R}_{j,l}} \cdot \bm{L}^{(\bm{R})}_{i,k;j,l}+
 \bm{\xi}^{(\bm{R})}_{i,k}(t) \\
 \label{langevin_equation_n_generic}
 & \frac{dN_{i,k}(t)}{dt} = - \sum_{j,l} L^{(N)}_{i,k;j,l} \frac{\partial
  \mathcal{F}_{\text{eff}}(\lbrace \bm{R}_{i,k} \rbrace,\lbrace N_{i,k}
  \rbrace, \lbrace Z_{i} \rbrace)}{\partial N_{i,k}}
 + k_{B} T \sum_{j,l} \frac{\partial L^{(N)}_{i,k;j,l}}{\partial N_{j,l}}  + \xi^{(N)}_{i,k}(t)
\end{align}
Here $\bm{L}^{(\bm{R})}_{i,k;j,l}$ and $L^{(N)}_{i,k;j,l}$ are the
mobility matrices, which may depend on
stochastic variables such as $\lbrace \bm{R}_{i,k} \rbrace$ or $\lbrace
N_{i,k} \rbrace$. (From the Onsager's reciprocal
relation, these mobility
matrices are symmetric.) $\mathcal{F}_{\text{eff}}(\lbrace \bm{R}_{i,k} \rbrace,\lbrace N_{i,k}
  \rbrace, \lbrace Z_{i} \rbrace)$ is the effective free energy of
the system, and the (generalized) forces are expressed as the derivatives of $\mathcal{F}_{\text{eff}}$.
$\bm{\xi}^{(\bm{R})}_{i,k}(t)$ and $\xi^{(N)}_{i,k}(t)$ are the
Gaussian random noises which satisfy the
the fluctuation-dissipation relations of the second kind.
They can be expressed as follows, by using Gaussian white noises
$\bm{w}_{i,k}^{\bm{R}}(t)$ and $w_{i,k}^{(N)}(t)$  (their statistical moments are given by eqs \eqref{w_r_fdr}-\eqref{w_r_n_fdr}).
\begin{align}
 & \label{xi_definition}
 \bm{\xi}^{(\bm{R})}_{i,k}(t) = \sqrt{2 k_{B} T} \sum_{j,l} \bm{B}_{i,k;j,l}^{(\bm{R})} \cdot \bm{w}_{j,l}^{(\bm{R})}(t),
 \qquad \xi^{(N)}_{i,k}(t) = \sqrt{2 k_{B} T}
 \sum_{j,l} B^{(N)}_{i,k;j,l} w_{j,l}^{(N)}(t) \\
 & \label{b_fdr}
 \sum_{m,n} \bm{B}_{i,k;m,n}^{(\bm{R})} \cdot
 [\bm{B}_{j,l;m,n}^{(\bm{R})}]^{\mathrm{T}} =
 \bm{L}_{i,k;j,l}^{(\bm{R})}, \qquad
 \sum_{m,n} B_{i,k;m,n}^{(N)} B_{j,l;m,n}^{(N)} =
 L_{i,k;j,l}^{(N)}
\end{align}
where $\bm{B}^{\mathrm{T}}$ represents the transposed matrix of $\bm{B}$.
In eqs \eqref{langevin_equation_r_generic} and
\eqref{langevin_equation_n_generic}, the stochastic terms are
interpreted following the Ito calculus\cite{Gardiner-book}.
The second terms in the right hand sides of eqs
\eqref{langevin_equation_r_generic} and
\eqref{langevin_equation_n_generic} are spurious (noise-induced) drift terms, which
are required to satisfy the detailed balance condition (for the Ito
form Langevin equations). The dynamics described by eqs
\eqref{langevin_equation_r_generic} is
\eqref{langevin_equation_n_generic} guaranteed to reach the equilibrium
state, of which statistics is simply determined by the effective free energy
$\mathcal{F}_{\text{eff}}$.

The PCN dynamic equation for
$\lbrace \bm{R}_{i,k} \rbrace$ (eq
\eqref{dynamic_equation_node_position}) seems to be similar to the
Langevin equation \eqref{langevin_equation_r_generic}.
Actually, it is straightforward to show that eq \eqref{dynamic_equation_node_position}
can be reduced to eq \eqref{langevin_equation_r_generic} by setting
\begin{equation}
 \label{mobility_node}
 \bm{L}^{(\bm{R})}_{i,k;j,l} = \frac{1}{\zeta} \delta_{ij} \delta_{kl} \bm{1}
\end{equation}
\begin{equation}
 \label{noise_node}
 \bm{\xi}^{(\bm{R})}_{i,k} = \sqrt{\frac{2 k_{B} T}{\zeta}} \bm{w}^{(\bm{R})}_{i,k}(t)
\end{equation}
and
\begin{equation}
 \label{effective_free_energy_pcn_partial}
 \mathcal{F}_{\text{eff}}(\lbrace \bm{R}_{i,k} \rbrace, \lbrace
  N_{i,k} \rbrace, \lbrace Z_{i} \rbrace)
  = \mathcal{F}_{\text{eff},0}(\lbrace \bm{R}_{i,k} \rbrace, \lbrace
  N_{i,k} \rbrace, \lbrace Z_{i} \rbrace)
  + \mathcal{F}_{\text{eff},1}(\lbrace N_{i,k}
  \rbrace,\lbrace Z_{i} \rbrace)
\end{equation}
\begin{equation}
 \label{effective_free_energy_pcn_partial_ideal}
 \mathcal{F}_{\text{eff},0}(\lbrace \bm{R}_{i,k} \rbrace, \lbrace
  N_{i,k} \rbrace, \lbrace Z_{i} \rbrace)
  \equiv \frac{3 k_{B} T} {2 b^{2}} \sum_{i} \sum_{k = 1}^{Z_{i}} \frac{(\bm{R}_{i,k} -
  \bm{R}_{i,k - 1})^{2}}{N_{i,k}}
\end{equation}
Here $\mathcal{F}_{\text{eff},0}(\lbrace \bm{R}_{i,k} \rbrace, \lbrace
N_{i,k} \rbrace, \lbrace Z_{i} \rbrace)$ is the free energy for ideal
linear springs, and
$\mathcal{F}_{\text{eff},1}(\lbrace N_{i,k} \rbrace,\lbrace Z_{i}
\rbrace)$ is the remaining contribution to the free energy which is
independent of $\lbrace \bm{R}_{i,k} \rbrace$.

On the other hand, the PCN dynamic equation for $\lbrace N_{i,k} \rbrace$ (eq
\eqref{dynamic_equation_segment_number} together with eq
\eqref{dynamic_equation_segment_number_flux}) is different from the standard Langevin equation
\eqref{langevin_equation_n_generic}.
The segment number flux in the PCN dynamic equation, eq
\eqref{dynamic_equation_segment_number_flux}, can be rewritten as
follows.
\begin{equation}
 \label{dynamic_equation_segment_number_flux_modified}
   J_{i,k}(t)
   = M_{i,k}
   \frac{3 k_{B} T}{\rho_{i,k} b^{2}}
   \left[
   \frac{|\bm{R}_{i,k + 1} - \bm{R}_{i,k}|}{N_{i,k + 1}}
   - \frac{|\bm{R}_{i,k} - \bm{R}_{i,k - 1}|}{N_{i,k}} \right] + \sqrt{2
   k_{B} T M_{i,k}} w_{i,k}(t)
\end{equation}
where we defined $M_{i,k}$ as
\begin{equation}
 \label{mobility_segment_flux}
 M_{i,k}(\lbrace \bm{R}_{i,k} \rbrace, \lbrace N_{i,k} \rbrace) \equiv \frac{2 \rho^{2}_{i,k}(\lbrace \bm{R}_{i,k} \rbrace, \lbrace N_{i,k} \rbrace)}{\zeta}
\end{equation}
Substituting eq \eqref{dynamic_equation_segment_number_flux_modified}
into \eqref{dynamic_equation_segment_number}, we have the following
expression for the dynamic equation. (For simplicity, we consider the case of $2 \le k \le Z_{i} - 1$ in eq \eqref{dynamic_equation_segment_number}.
The generalization to $k = 1, Z_{i}$ is
straightforward.)
\begin{equation}
 \label{dynamic_equation_segment_number_modified}
\begin{split}
   \frac{dN_{i,k}(t)}{dt}
 = & - M_{i,k} \frac{3 k_{B} T}{\rho_{i,k} b^{2}} \bigg[
   \frac{|\bm{R}_{i,k + 1} - \bm{R}_{i,k}|}{N_{i,k + 1}}
 - \frac{|\bm{R}_{i,k} - \bm{R}_{i,k -
 1}|}{N_{i,k}} \bigg]
 \\
 & + M_{i,k - 1} \frac{3 k_{B} T}{\rho_{i,k - 1} b^{2}} \bigg[
\frac{|\bm{R}_{i,k} - \bm{R}_{i,k - 1}|}{N_{i,k}}
   - 
 \frac{|\bm{R}_{i,k - 1} - \bm{R}_{i,k - 2}|}{N_{i,k - 1}} \bigg] \\
 &- \sqrt{2
   k_{B} T M_{i,k}} w_{i,k}(t) + \sqrt{2
   k_{B} T M_{i,k - 1}} w_{i,k - 1}(t)
\end{split}
\end{equation}
Comparing the stochastic terms in eqs \eqref{dynamic_equation_segment_number_modified} and
\eqref{langevin_equation_n_generic}, we find that the mobility matrix
$L_{i,k;j,l}^{(N)}$ can be expressed by using $M_{i,k}$ as
\begin{equation}
 \label{mobility_segment_transport}
 L_{i,k;j,l}^{(N)} =
 \begin{cases}
  M_{i,k} + M_{i,k - 1} & (i = j, l = k) \\
  - M_{i,k - 1} & (i = j, l = k - 1) \\
  - M_{i,k} & (i = j, l = k + 1) \\
  0 & (\text{otherwise})
 \end{cases}
\end{equation}
We also find that the deterministic terms in eq
\eqref{dynamic_equation_segment_number_modified} {\em can not be reduced to
the variational form} in eq \eqref{langevin_equation_n_generic}.
Namely, the PCN model does not satisfy the detailed balance condition
and thus it {\em does not have the thermal equilibrium state}. The
steady state in the PCN model (in absence of external flow field)
corresponds to a nonequilibrium steady state, where the energy input and
the energy dissipation is balanced \cite{Sekimoto-book}.

Nevertheless, PCN simulations by the original dynamic equations can reproduce some equilibrium properties
of entangled polymers
reasonably\cite{Masubuchi-Uneyama-Watanabe-Ianniruberto-Greco-Marrucci-2010}.
This implies that, although
the PCN model is not detailed-balanced and does
not have the equilibrium state, it works as a good approximation of a
statistical mechanically sound, detailed-balanced dynamical model.
Keeping this in mind, we consider to remedy the PCN model to satisfy the
detailed balance condition in a heuristic way.
From eqs \eqref{dynamic_equation_segment_number_modified} and
\eqref{langevin_equation_n_generic},
the variational form for the force can be realized if
the following relation holds.
\begin{equation}
 \label{dynamic_equation_segment_number_variational_condition}
  \frac{3 k_{B} T}{\rho_{i,k} b^{2}}
  \bigg[
 \frac{|\bm{R}_{i,k} - \bm{R}_{i,k - 1}|}{N_{i,k}}
  - \frac{|\bm{R}_{i,k + 1} - \bm{R}_{i,k}|}{N_{i,k + 1}} \bigg] =
 \bigg( \frac{\partial}{\partial N_{i,k + 1}}
 - \frac{\partial}{\partial N_{i,k}} \bigg)
 \mathcal{F}_{\text{eff}}(\lbrace \bm{R}_{i,k} \rbrace,\lbrace N_{i,k}
 \rbrace, \lbrace Z_{i} \rbrace)
\end{equation}
Clearly, eq
\eqref{dynamic_equation_segment_number_variational_condition} does not
hold if we use eq \eqref{line_density_definition} as the definition of $\rho_{i,k}$.
Conversely, if we
do not use eq \eqref{line_density_definition} as the definition of
$\rho_{i,k}$, it becomes possible to satisfy eq
\eqref{dynamic_equation_segment_number_variational_condition}.
Besides, there is no special reason to employ the arithmetic average form for
$\rho_{i,k}$. (Different models such as a constant value model were also
proposed and utilized\cite{Masubuchi-Takimoto-Koyama-Ianniruberto-Greco-Marrucci-2001}.)
Thus, here we {\em define the segment density along the chain
$\rho_{i,k}$ to recover the variational form}.
\begin{equation}
 \label{line_density_definition_alternative}
  \rho_{i,k}(\lbrace \bm{R}_{i,k} \rbrace, \lbrace N_{i,k} \rbrace) \equiv 2 \left[ 
  \frac{|\bm{R}_{i,k + 1} - \bm{R}_{i,k}|}{N_{i,k + 1}}
   + \frac{|\bm{R}_{i,k} - \bm{R}_{i,k - 1}|}{N_{i,k}}  \right]^{-1}
\end{equation}
The right hand side of eq \eqref{line_density_definition_alternative} is
the {\em harmonic average} of $N_{i,k + 1} / |\bm{R}_{i,k + 1} -
\bm{R}_{i,k}|$ and $N_{i,k} / |\bm{R}_{i,k} - \bm{R}_{i,k - 1}|$,
whereas the right hand side of eq \eqref{line_density_definition} is the
{\em arithmetic average} of them.
By using the new definition \eqref{line_density_definition_alternative},
it is straightforward to show that the condition
\eqref{dynamic_equation_segment_number_variational_condition} is satisfied and
$\mathcal{F}_{\text{eff},1} = 0$.

Now the PCN dynamic equation for $\lbrace N_{i,k} \rbrace$
(eq \eqref{dynamic_equation_segment_number_modified}) can be
rewritten as follows.
\begin{equation}
 \label{dynamic_equation_segment_number_modified_with_new_linear_density}
   \frac{dN_{i,k}(t)}{dt}
 = - \sum_{j,l} L^{(N)}_{i,k;j,l} \frac{\partial
 \mathcal{F}_{\text{eff,0}}(\lbrace \bm{R}_{i,k} \rbrace, \lbrace
 N_{i,k} \rbrace, \lbrace Z_{i} \rbrace)}{\partial N_{j,l}}
 + \xi_{i,k}^{(N)}(t)
\end{equation}
with
\begin{equation}
 \xi_{i,k}^{(N)}(t) = - \sqrt{2
   k_{B} T M_{i,k}} w_{i,k}^{(N)}(t) + \sqrt{2
   k_{B} T M_{i,k - 1}} w_{i,k - 1}^{(N)}(t)
\end{equation}
Although eq
\eqref{dynamic_equation_segment_number_modified_with_new_linear_density}
is similar to eq \eqref{langevin_equation_n_generic}, the spurious
drift term (the second term in the right hand side of eq
\eqref{langevin_equation_n_generic}) is missing in eq 
\eqref{dynamic_equation_segment_number_modified_with_new_linear_density}.
Therefore, in addition to employ a new definition of $\rho_{i,k}$ (eq \eqref{line_density_definition_alternative}), we
{\em add the spurious drift term to the segment number flux equation} (eq \eqref{dynamic_equation_segment_number_flux}).
We modify the definition for the segment flux (eq
\eqref{dynamic_equation_segment_number_flux}) as
\begin{equation}
 \label{dynamic_equation_segment_number_flux_alternative}
   \frac{\zeta}{2 \rho_{i,k}}
   J_{i,k}(t)
   \equiv
    \displaystyle \frac{3 k_{B} T}{b^{2}}
   \left[ \frac{|\bm{R}_{i,k + 1} - \bm{R}_{i,k}|}{N_{i,k + 1}}
   - \frac{|\bm{R}_{i,k} - \bm{R}_{i,k - 1}|}{N_{i,k}} \right]
    + v_{i,k}^{(N)}
   + \sqrt{k_{B} T \zeta} w^{(N)}_{i,k}(t)
\end{equation}
where $v_{i,k}^{(N)}$ is the spurious drift velocity defined as
\begin{equation}
 \label{extra_drift_n_flux_definition}
 \begin{split}
 v^{(N)}_{i,k}
  & \equiv
  2 k_{B} T
  \left( \frac{\partial}{\partial N_{i,k + 1}} - \frac{\partial}{\partial N_{i,k}}
    \right) \rho_{i,k} \\
  & = 4 k_{B} T
  \frac{N^{2}_{i,k} |\bm{R}_{i,k + 1} - \bm{R}_{i,k}|
  - N^{2}_{i,k + 1} |\bm{R}_{i,k} - \bm{R}_{i,k - 1}|}
  {( N_{i,k} |\bm{R}_{i,k + 1} - \bm{R}_{i,k}|
  + N_{i,k + 1} |\bm{R}_{i,k} - \bm{R}_{i,k - 1}| )^{2}}
 \end{split}
\end{equation}
The modified PCN dynamic equation for $\lbrace N_{i,k} \rbrace$ (eq
\eqref{dynamic_equation_segment_number} together with eqs
\eqref{line_density_definition_alternative},
\eqref{dynamic_equation_segment_number_flux_alternative}, and
\eqref{extra_drift_n_flux_definition}) reduces to eq
\eqref{langevin_equation_n_generic}. Thus the modified PCN dynamic
equations satisfy the detailed balance condition.
It should be emphasized here that the noise term in eq
\eqref{langevin_equation_n_generic} is the {\em multiplicative noise} because
$L_{i,k;j,l}^{(N)}$ depends on $\lbrace N_{i,k} \rbrace$ (through
$M_{i,k}$ and $\rho_{i,k}$).
Although the spurious drift term may not be intuitive, it naturally
arises as a property of the multiplicative noise
\cite{Risken-book,Kindt-Briels-2007}, and is required to satisfy the
detailed balance condition.

Before we proceed to detailed analysis, here we shortly comment on the
relation of our model  to the GENERIC framework in the nonequilibrium
thermodynamics \cite{Beris-Edwards-book,Ottinger-1998}.
In our modified PCN dynamic equations,
the thermodynamic forces are coupled to the symmetric mobility matrices.
Such dynamic equations can be straightforwardly mapped onto the GENERIC
framework.
The mobility matrices and the
effective free energy in our model correspond to the friction matrices
and the entropy (with negative sign), respectively. They form the
irreversible, dissipative bracket parts. The reversible, Poisson bracket
parts do not exist in the modified PCN dynamic equations.
Most of
important properties in the nonequilibrium thermodynamics, such as the
positivity of the entropy production, are 
automatically reproduced in the modified PCN dynamic equations.
(Notice that, however, this does not mean that the modified
PCN model is fully consistent with the GENERIC framework and the
nonequilibrium thermodynamics. The repulsive
interaction or the network reconstruction process may be inconsistent
with them.)

\subsection{Effective Free Energy}

The modified PCN dynamic equations introduced in the previous subsection
reduces to the detailed-balanced Langevin equation.
The equilibrium state of the modified PCN model is determined by
the effective free energy $\mathcal{F}_{\text{eff}}$.
The effective free energy of the modified PCN model can
be simply expressed as
\begin{equation}
 \label{effective_free_energy_pcn}
  \mathcal{F}_{\text{eff}}(\lbrace \bm{R}_{i,k} \rbrace, \lbrace
  N_{i,k} \rbrace, \lbrace Z_{i} \rbrace)
   = \mathcal{F}_{\text{eff},0}(\lbrace \bm{R}_{i,k} \rbrace, \lbrace
  N_{i,k} \rbrace, \lbrace Z_{i} \rbrace)
   = \frac{3 k_{B} T} {2 b^{2}} \sum_{i} \sum_{k = 1}^{Z_{i}} \frac{(\bm{R}_{i,k} -
  \bm{R}_{i,k - 1})^{2}}{N_{i,k}}
\end{equation}
Notice that, the effective free energy \eqref{effective_free_energy_pcn}
is different from the free energy of ideal Gaussian
chains. The free energy of Gaussian chains is expressed as
\begin{equation}
 \label{free_energy_gaussian_chains}
  \mathcal{F}_{0}(\lbrace \bm{R}_{i,k} \rbrace, \lbrace N_{i,k} \rbrace,
  \lbrace Z_{i} \rbrace)
  \equiv \frac{3 k_{B} T} {2 b^{2}} \sum_{i} \sum_{k = 1}^{Z_{i}} \frac{(\bm{R}_{k} -
  \bm{R}_{k - 1})^{2}}{N_{k}}
  + \frac{3 k_{B} T}{2} \sum_{i} \sum_{k = 1}^{Z_{i}} \ln N_{k}
\end{equation}
Comparing eqs \eqref{effective_free_energy_pcn} and
\eqref{free_energy_gaussian_chains}, we find that the logarithm term is
missing in the PCN effective free energy.

We can rewrite the PCN effective free energy
\eqref{effective_free_energy_pcn} as follows, by using the free
energy of Gaussian chains \eqref{free_energy_gaussian_chains}.
\begin{align}
 & \label{effective_free_energy_pcn_modified}
  \mathcal{F}_{\text{eff}}(\lbrace \bm{R}_{i,k} \rbrace, \lbrace
  N_{i,k} \rbrace, \lbrace Z_{i} \rbrace)
   =  \mathcal{F}_{0}(\lbrace \bm{R}_{i,k} \rbrace, \lbrace
  N_{i,k} \rbrace, \lbrace Z_{i} \rbrace)
 + \sum_{i} \sum_{k = 1}^{Z_{i}} \tilde{U}_{\text{slip-link}}(N_{k}) \\
 & \label{effective_potential_pcn_sliplink}
 \tilde{U}_{\text{slip-link}}(n) \equiv - \frac{3 k_{B} T}{2} \ln n
\end{align}
Here $\tilde{U}_{\text{slip-link}}(n)$ represents the effective
interaction between neighboring slip-links.
(Strictly speaking,
$\tilde{U}_{\text{slip-link}}$ also represents the effective interaction
potential between a slip-link and a chain end. In the followings, we assume
that the interaction potentials between the slip-links and between a
slip-link and a chain end are always the same.)
We note that $\tilde{U}_{\text{slip-link}}(n)$ depends only on the number of
segments between slip-links. This means that the effective interaction
between slip-links is determined by the chemical distance, not by the
spatial distance.
Eq \eqref{effective_potential_pcn_sliplink} is monotonically decreasing
function of $n$ and thus the interaction between slip-links is purely repulsive.
In addition, the effective potential diverges at the
limit of $n \to 0$, and thus slip-links strongly repel
each other at the short range (at the short chemical distance).

The second term in the right hand side of eq
\eqref{effective_free_energy_pcn_modified} can be interpreted as
the total repulsive interaction energy of slip-links.
In this sense,
we can say that the PCN model is based on the Gaussian chain and repulsive
slip-links.
It is worth mentioning that such an effective repulsion between
slip-links is
already suggested by the CReTA primitive path extraction for Monte Carlo
simulation data \cite{Tzoumanekas-Theodorou-2006}.

\section{Equilibrium Probability Distributions by Single Chain Approximation}
\label{equilibrium_probability_distributions_for_single_primitive_chain_model}

\subsection{Single Chain Approximation}

In the previous section, the effective free energy for the modified PCN model
\eqref{effective_free_energy_pcn} is shown to be different from the
free energy for Gaussian chains \eqref{free_energy_gaussian_chains}, due to the repulsive
interaction between slip-links.
This means that several equilibrium statistical properties of the PCN
model are qualitatively different from ones of the Gaussian chains with
slip-links without any direction interactions.
However, unfortunately we cannot calculate the equilibrium statistics of PCN
analytically from the effective free energy.
(It is practically impossible to calculate the statistical weight for
strongly slip-linked multi chain systems.)
In this section, we calculate
the equilibrium statistical properties by using the mean field type single
chain approximation.
(We notice that, it is not clear whether the single chain approximation
employed in this work is really reasonable for the PCN model or not.
We employ the simplest approximation to make the expressions simple and
analytically tractable.)

Schieber \cite{Schieber-2003} proposed a single chain model with
slip-links in which slip-links behave as a sort
of grand canonical gas particles on a polymer chain.
In his model, the number of slip-linked subchains is controlled by the
effective chemical potential.
There is no direct interaction between slip-links and the polymer chain obeys
the Gaussian statistics. Namely, slip-links are placed on the target
polymer chain and they do not directly interact each other via a
potential (this model corresponds to a sort of
ideal model).
In this work, we call his model as ``the
single chain non-interacting slip-link model'' or simply as ``the
non-interacting slip-link model''.
(The expression ``non-interacting'' may sound somehow
unnatural, because the slip-links constrain the polymer chain and this
constraint effect can be interpreted as a sort of interaction.
Nonetheless, in this work we employ
this expression because in the followings we consider another model in which
slip-links directly ``interact'' each other.)
The statistics of the single chain
non-interacting slip-link model is briefly summarized in Appendix
\ref{equilibrium_distribution_functions_for_entangled_signle_gaussian_chain}.

Here we follow Schieber's idea and consider the single chain version of
the modified PCN model.
We may call this as the ``the single chain repulsive slip-link model'' or
``the repulsive slip-link model'' in the followings. This is because in
our model, neighboring slip-links interact each other via the repulsive potential
(eq \eqref{effective_potential_pcn_sliplink}).
In the single chain repulsive slip-link model, the state of the system is
expressed by the node positions $\lbrace \bm{R}_{k} \rbrace$ ($k$
represents the $k$-th segment on the chain), the
numbers of segments $\lbrace N_{k} \rbrace$, and the number of subchains
$Z$. (Under the mean field type single chain approximation, the connectivity map is
not required.)

The effective free energy for a single Gaussian chain with repulsive
slip-links can be expressed
as the single chain version of eq
\eqref{effective_free_energy_pcn}.
\begin{equation}
 \label{effective_free_energy_single_chain}
  \mathcal{F}_{\text{eff}}(\lbrace \bm{R}_{k} \rbrace, \lbrace N_{k} \rbrace, Z)
  = \frac{3 k_{B} T} {2 b^{2}} \sum_{k = 1}^{Z} \frac{(\bm{R}_{k} -
  \bm{R}_{k - 1})^{2}}{N_{k}}
\end{equation}
Introducing the effective chemical potential for subchains, $\epsilon$, we define
the following effective grand potential (grand canonical thermodynamic potential) for the
repulsive slip-link model.
\begin{equation}
 \label{effective_grand_potential_single_chain}
  \mathcal{J}_{\text{eff}}(\lbrace \bm{R}_{k} \rbrace, \lbrace N_{k} \rbrace, Z)
  \equiv \mathcal{F}_{\text{eff}}(\lbrace \bm{R}_{k} \rbrace, \lbrace N_{k} \rbrace, Z)
   - \epsilon Z
\end{equation}
The effective chemical potential $\epsilon$ is determined from the
following condition for the equilibrium average number of subchains
\begin{equation}
 \label{condition_for_effective_chemical_potential}
 \langle Z \rangle_{\text{eq}} \equiv Z_{0} = \frac{N}{N_{0}}
\end{equation}
where $\langle \dots \rangle_{\text{eq}}$ represents the equilibrium
statistical average calculated by the equilibrium probability
distribution. Here $N$ is the number of segments in a chain and 
$N_{0}$ corresponds to the equilibrium average number of
segments in a subchain.

All the equilibrium statistical properties can be calculated by the
effective grand potential \eqref{effective_grand_potential_single_chain}.
The equilibrium statistical probability distribution is given as the
standard Boltzmann distribution.
\begin{equation}
 \label{equilibrium_distirbution_pcn}
 P_{\text{eq}}(\lbrace \bm{R}_{k} \rbrace, \lbrace N_{k} \rbrace, Z) =
  \frac{1}{\Xi \Lambda^{3 (Z + 1)} N^{Z - 1}}
  \delta \bigg( N - \sum_{k = 1}^{Z} N_{k} \bigg)
  \exp \left[ - \frac{\mathcal{J}_{\text{eff}}(\lbrace \bm{R}_{k} \rbrace,
	\lbrace N_{k} \rbrace, Z)}{k_{B} T} \right]
\end{equation}
where $\Lambda$ is the thermal de Broglie wave length and $\Xi$ is the
grand partition function.
(The subscript ``eq'' in eq
\eqref{equilibrium_distirbution_pcn} refers to the equilibrium state.)
The thermal de Broglie wave length was introduced to make the grand
partition function dimensionless.
The delta function in eq \eqref{equilibrium_distirbution_pcn} comes
from the constraint that the total number of segments in a chain is
constant. The
grand partition function $\Xi$ is defined as
\begin{equation}
 \label{partition_function_single_chain}
   \Xi
   \equiv \sum_{Z = 1}^{\infty} \frac{1}{\Lambda^{3 (Z + 1)} N^{Z - 1}}
   \int d\lbrace \bm{R}_{k} \rbrace d\lbrace N_{k} \rbrace \,
  \delta \bigg( N - \sum_{k = 1}^{Z} N_{k} \bigg)
  \exp \left[ - \frac{\mathcal{J}_{\text{eff}}(\lbrace \bm{R}_{k} \rbrace,
	\lbrace N_{k} \rbrace, Z)}{k_{B} T} \right]
\end{equation}
Here we introduced the shorthand notation for the integrals over
$\lbrace \bm{R}_{k} \rbrace$ and $\lbrace N_{k} \rbrace$.
\begin{equation}
   \int d\lbrace \bm{R}_{k} \rbrace d\lbrace N_{k} \rbrace \,
    \equiv
  \prod_{k = 0}^{Z} \int d\bm{R}_{k}
  \prod_{k = 1}^{Z} \int_{0}^{N}
  dN_{k}
\end{equation}

By calculating the integrals over $\lbrace \bm{R}_{k} \rbrace$, eq \eqref{partition_function_single_chain}
becomes as follows.
\begin{equation}
 \label{partition_function_single_chain_modified}
  \Xi = \frac{\mathcal{V}}{\Lambda^{3}} \sum_{Z = 1}^{\infty}
  \bigg[ \left(\frac{2 \pi N b^{2}}{3 \Lambda^{2}} \right)^{3 /2} e^{\epsilon /
  k_{B} T} \bigg]^{Z}
  \frac{1}{N^{5 Z / 2 - 1}} \int
  d\lbrace N_{k} \rbrace \,
  \delta \bigg( N - \sum_{k = 1}^{Z} N_{k} \bigg)
  \prod_{k = 1}^{Z} N_{k}^{3/2}
\end{equation}
where $\mathcal{V}$ is the volume of the system.
The integrals over $\lbrace N_{k} \rbrace$ in eq
\eqref{partition_function_single_chain_modified} can be analytically
calculated as
\begin{equation}
 \label{n_integral_modified}
   \frac{1}{N^{5 Z / 2 - 1}} \int
   d\lbrace N_{k} \rbrace \,
   \delta \bigg( N - \sum_{k = 1}^{Z} N_{k} \bigg)
   \prod_{k = 1}^{Z} N_{k}^{3/2}
   = \frac{(3 \sqrt{\pi} / 4)^Z}{\Gamma(5 Z / 2)}
\end{equation}
where $\Gamma(x)$ is the gamma
function \cite{Abramowitz-Stegun-book}.
The detailed calculation is described in Appendix \ref{integrals_over_n_k}.
By substituting eq \eqref{n_integral_modified} into eq \eqref{partition_function_single_chain_modified},
finally we have the following expression.
\begin{equation}
 \label{partition_function_single_chain_modified2}
  \begin{split}
   \Xi
   & = \frac{\mathcal{V}}{\Lambda^{3}} \sum_{Z = 1}^{\infty}
   \frac{1}{\Gamma(5
   Z / 2)} \bigg[  \frac{3 \pi^{2}}{4} \left(\frac{2 N b^{2}}{3 \Lambda^{2}} \right)^{3 /2} e^{\epsilon /
   k_{B} T} \bigg]^{Z} 
   = \frac{\mathcal{V}}{\Lambda^{3}} \xi E_{5/2,5/2}(\xi)
  \end{split}
\end{equation}
Here we defined the
dimensionless effective fugacity $\xi$ as
\begin{equation}
 \xi \equiv \frac{3 \pi^{2}}{4} \left(\frac{2 N b^{2}}{3
					 \Lambda^{2}} \right)^{3 /2}
 e^{\epsilon /  k_{B} T}
\end{equation}
$\xi$ is determined to satisfy the condition for $\langle Z
\rangle_{\text{eq}}$ (eq \eqref{condition_for_effective_chemical_potential}).
$E_{\alpha,\beta}(x)$ in the last line of eq
\eqref{partition_function_single_chain_modified2} is the (generalized)
Mittag-Leffler function \cite{Weisstein-mathworld} defined as
\begin{equation}
 \label{mittag_leffler_function}
 E_{\alpha,\beta}(x) \equiv \sum_{n = 0}^{\infty}
  \frac{x^{n}}{\Gamma(\alpha n + \beta)}
\end{equation}
Unfortunately, the Mittag-Leffler
function is not easy to handle.
Therefore, in the following subsections we derive simple and approximate
expressions rather than exact expressions.

\subsection{Slip-Linked Subchain Number Distribution}
\label{sliplinkd_subchain_number_distribution}

First we calculate the slip-linked subchain number distribution.
The distribution function of the subchain number $Z$ can
be expressed as follows.
\begin{equation}
 \label{sliplinked_distribution_single_chain_definition}
 P_{\text{eq}}(Z) =
 \int d\lbrace \bm{R}_{k} \rbrace d\lbrace N_{k} \rbrace \,
 P_{\text{eq}}(\lbrace \bm{R}_{k} \rbrace,\lbrace N_{k} \rbrace, Z)
\end{equation}
By integrating eq
\eqref{sliplinked_distribution_single_chain_definition} over $\lbrace \bm{R}_{k} \rbrace$ and $\lbrace N_{k}
\rbrace$, we have the following expression.
\begin{equation}
 \label{sliplinked_distribution_single_chain_modified}
 P_{\text{eq}}(Z)
   = \frac{\xi^{Z - 1}}{\Gamma(5 Z / 2) E_{5/2,5/2}(\xi)}
\end{equation}
Eq \eqref{sliplinked_distribution_single_chain_modified} can not be
expressed by elementary functions.
This is in contrast to the non-interacting slip-link model, of which
subchain number distribution is simply given as
a Poisson distribution (eq
\eqref{sliplinked_distribution_single_gaussian_chain_modified} in
Appendix \ref{equilibrium_distribution_functions_for_entangled_signle_gaussian_chain}).

\begin{figure}[h!]
 \centering
 {\includegraphics[width=0.95\linewidth,clip]{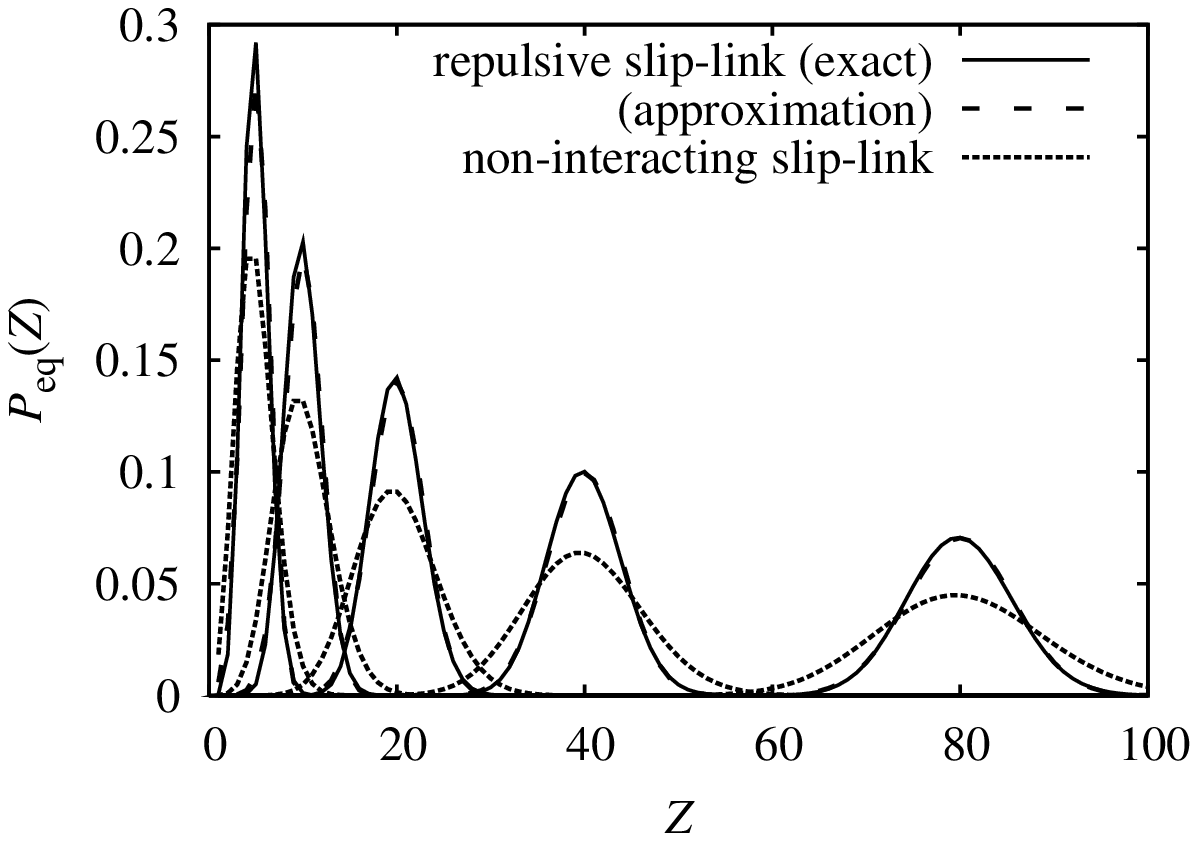}}
 \caption{}
 \label{sliplinked_number_distribution_function_single_chain_graph}
\end{figure}

We can utilize the saddle point approximation to obtain a simple
approximate form for eq
\eqref{sliplinked_distribution_single_chain_modified}.
(See Appendix \ref{saddle_point_approximation_for_p_z}.)
\begin{equation}
\label{sliplinked_distribution_single_chain_approximated}
 P_{\text{eq}}(Z) \approx \sqrt{\frac{5 Z_{0} + 1}{4 \pi Z_{0}^{2}} }
  \exp \left[ - 
              \frac{5 Z_{0} + 1}{4 Z_{0}^{2}}
             (Z - Z_{0})^{2} \right]
\end{equation}
Figure
\ref{sliplinked_number_distribution_function_single_chain_graph} shows
the subchain number distribution functions for various values of $Z_{0}$
by eq \eqref{sliplinked_distribution_single_chain_modified} (exact)
and by eq \eqref{sliplinked_distribution_single_chain_approximated} (approximation).
For comparison, the distribution function of the non-interaction
slip-link model (eq
\eqref{sliplinked_distribution_single_gaussian_chain_modified} in
Appendix \ref{equilibrium_distribution_functions_for_entangled_signle_gaussian_chain}) is also shown in Figure \ref{sliplinked_number_distribution_function_single_chain_graph}.
We find that the approximate form
\eqref{sliplinked_distribution_single_chain_approximated} works very
well even for relatively small $Z_{0}$ such as $Z_{0} = 5$.

\begin{figure}[c!]
 \centering
 {\includegraphics[width=0.95\linewidth,clip]{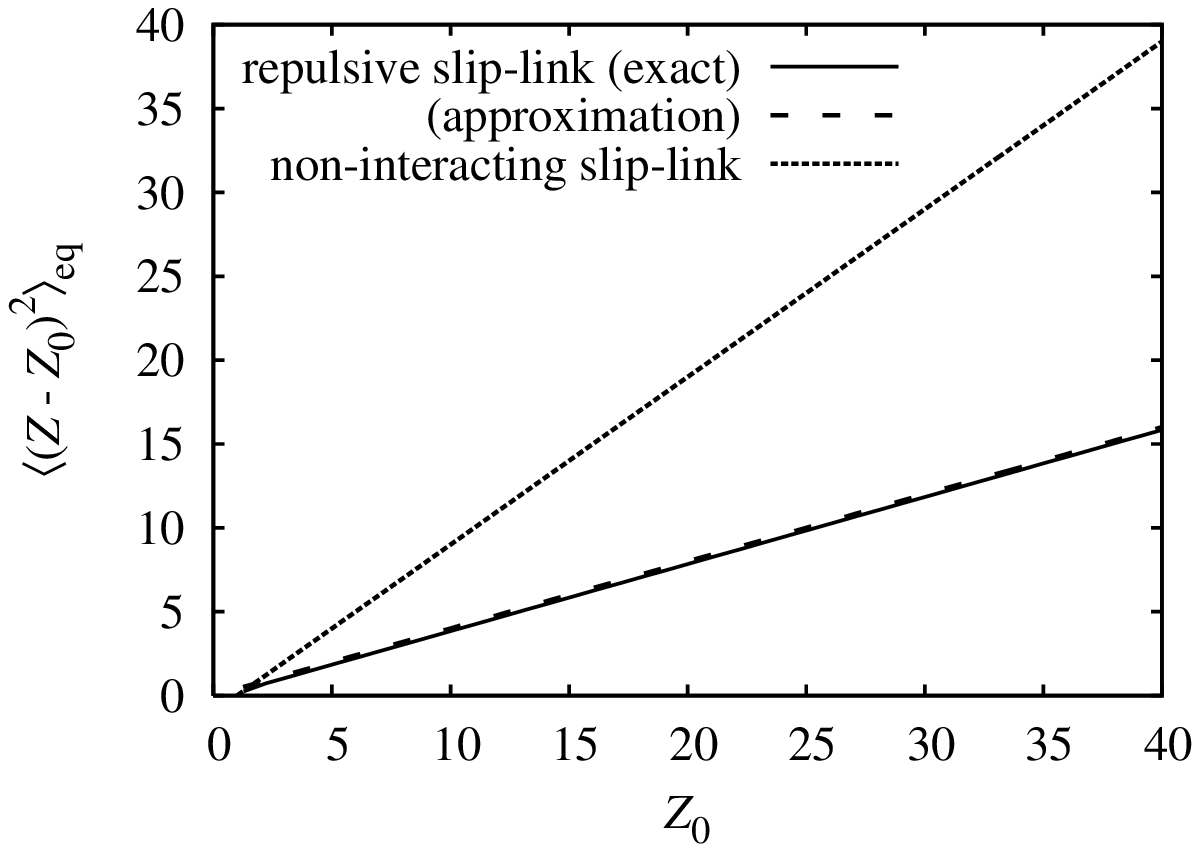}}
 \caption{}
 \label{average_and_variance_of_sliplinked_number_graph}
\end{figure}

The relation between the average and variance of $Z$ is easily obtained
from eq
\eqref{sliplinked_distribution_single_chain_approximated}.
\begin{equation}
 \label{sliplinked_distribution_single_chain_approximated_variance}
 \langle (Z - Z_{0})^{2} \rangle_{\text{eq}} \approx
 \frac{2 Z_{0}^{2}}{5 Z_{0} + 1} \approx \frac{2}{5} Z_{0} \qquad
 (\text{for $Z_{0} \gg 1$})
\end{equation}
Figure \ref{average_and_variance_of_sliplinked_number_graph} shows the
relation between $\langle Z \rangle_{\text{eq}} = Z_{0}$ and $\langle (Z
- Z_{0})^{2} \rangle_{\text{eq}}$, calculated by the exact and
approximate forms. Again we find that the approximate form
\eqref{sliplinked_distribution_single_chain_approximated} works well
even for relatively small $Z_{0}$.
For sufficiently large $Z_{0}$, we find that the variance of the repulsive slip-link
model is smaller than one for non-interacting slip-link model by the factor $2/5$ (see
eq \eqref{variance_z_single_gaussian_chain} in Appendix
\ref{equilibrium_distribution_functions_for_entangled_signle_gaussian_chain}).
The distribution of $Z$ of the repulsive slip-link model is sharper than one of the non-interacting
slip-link model, as shown in Figure \ref{sliplinked_number_distribution_function_single_chain_graph}.

\subsection{Segment Number Distribution}
\label{segment_number_distribution}

Second we calculate the distribution function of the segment number in a
subchain.
We express the number of segments as $n$. The segment number
distribution function $P_{\text{eq}}(n)$ is expressed as follows.
\begin{equation}
 \label{segment_distribution_single_chain_definition}
   P_{\text{eq}}(n) = \sum_{Z = 1}^{\infty}
   \int d\lbrace \bm{R}_{k} \rbrace
   d\lbrace N_{k} \rbrace \, \bigg[ \frac{1}{Z} \sum_{l = 1}^{Z} \delta(n - N_{l}) \bigg]
   P_{\text{eq}}(\lbrace \bm{R}_{k} \rbrace,
   \lbrace N_{k} \rbrace, Z)
\end{equation}
By calculating the integrals over $\lbrace \bm{R}_{k} \rbrace$ in eq
\eqref{segment_distribution_single_chain_definition}, we
have
\begin{equation}
 \label{segment_distribution_single_chain_definition_modified}
   P_{\text{eq}}(n)
   =
   \frac{\mathcal{V}}{\Xi \Lambda^{3} N} \sum_{Z = 1}^{\infty}
   \frac{\xi^{Z}}{N^{5 Z / 2 - 2}} \int
   d\lbrace N_{k} \rbrace \, \bigg[ \frac{1}{Z} \sum_{l = 1}^{Z} \delta(n - N_{l})
   \bigg]
   \delta \bigg( N - \sum_{k = 1}^{Z} N_{k} \bigg)
  \prod_{k = 1}^{Z} N_{k}^{3/2}
\end{equation}
Integrals over $\lbrace N_{k} \rbrace$ in eq
\eqref{segment_distribution_single_chain_definition_modified} can be
calculated analytically. For $Z \ge 2$ we have
\begin{equation}
 \label{n_integral_with_constraint_modified}
  \begin{split}
   & \frac{1}{N^{5 Z / 2 - 2}} \int
   d\lbrace N_{k} \rbrace \, \bigg[ \frac{1}{Z} \sum_{l = 1}^{Z} \delta(n - N_{l})
   \bigg]
   \delta \bigg( N - \sum_{k = 1}^{Z} N_{k} \bigg)
  \prod_{k = 1}^{Z} N_{k}^{3/2} \\
   & = 
   \left( \frac{n}{N} \right)^{3/2}
   \left( 1 - \frac{n}{N} \right)^{5 (Z - 1) / 2 - 1}
   \frac{(3 \sqrt{\pi} / 4)^{Z - 1}}{\Gamma(5 (Z - 1) / 2)}
  \end{split}
\end{equation}
(see Appendix \ref{integrals_over_n_k}) and for $Z = 1$ we have
\begin{equation}
   \frac{1}{N^{1/2}} \int_{0}^{N}
   dN_{1} \, \delta(n - N_{1})
   \delta ( N - N_{1})
  N_{1}^{3/2}
   = \delta\left( \frac{n}{N} - 1 \right)
\end{equation}
Thus eq \eqref{segment_distribution_single_chain_definition_modified}
can be rewritten as follows.
\begin{equation}
 \label{segment_distribution_single_chain_definition_modified2}
  \begin{split}
   P_{\text{eq}}(n)
   & = \frac{4}{3 \sqrt{\pi} N E_{5/2,5/2}(\xi)}
\bigg[ \delta\left( \frac{n}{N} - 1 \right) + \sum_{Z = 2}^{\infty}
   \left( \frac{n}{N} \right)^{3/2}
   \left( 1 - \frac{n}{N} \right)^{5 (Z - 1) / 2 - 1}
   \frac{\xi^{Z - 1}}{\Gamma(5 (Z - 1) / 2)} \bigg]
  \end{split}
\end{equation}

\begin{figure}[c!]
 \centering
 {\includegraphics[width=0.95\linewidth,clip]{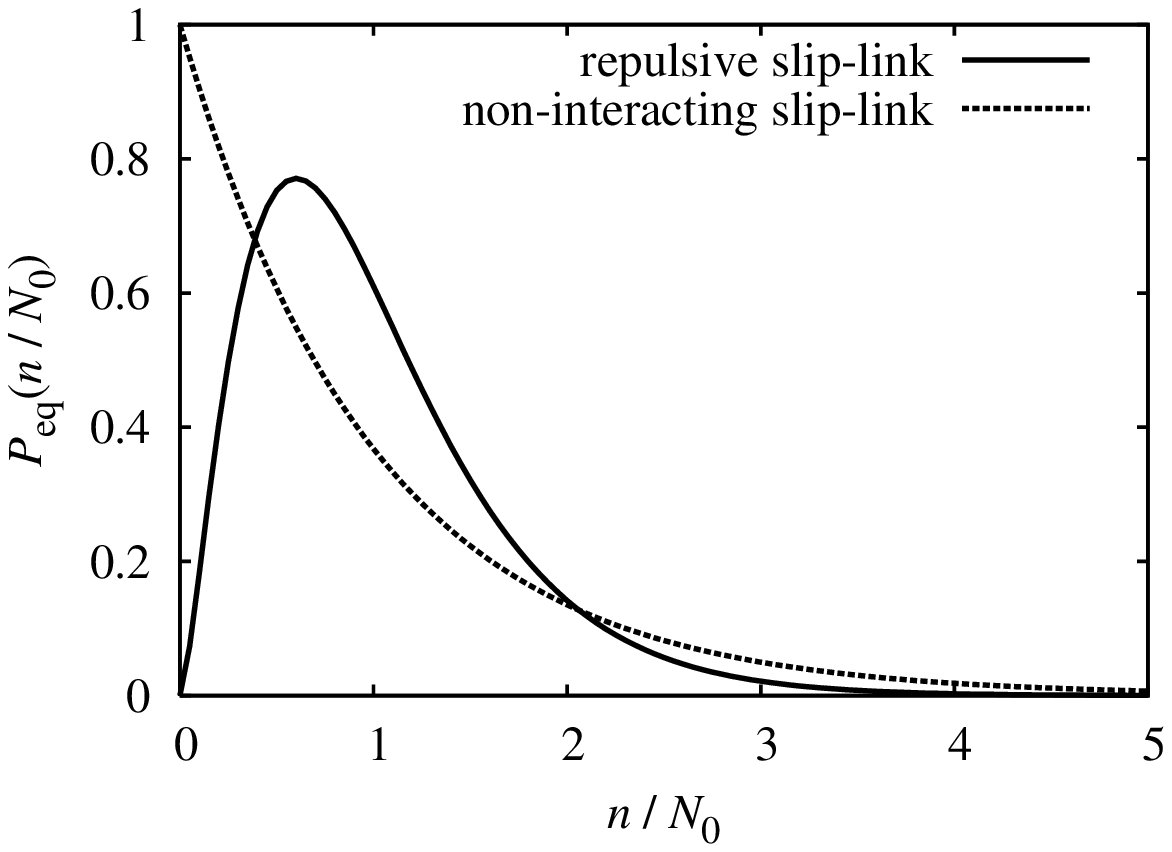}}
 \caption{}
 \label{segment_number_distribution_function_single_chain_graph}
\end{figure}

As before, we can obtain a simple approximate expression by using the
saddle point approximation. (See Appendix
\ref{saddle_point_approximation_for_p_n}.)  For sufficiently large
$Z_{0}$, eq \eqref{n_integral_with_constraint_modified} is approximated as
\begin{equation}
 \label{segment_distribution_single_chain_definition_approximated_final}
  P_{\text{eq}}(n)
  \approx
   \frac{25}{6} \sqrt{\frac{10}{\pi}}
   \frac{n^{3/2}}{N_{0}^{5/2}}
   \exp \left( - \frac{5 n}{2 N_{0}} \right)
\end{equation}
Eq \eqref{segment_distribution_single_chain_definition_approximated_final} is qualitatively different from
one of the single chain non-interacting slip-link model (eq
\eqref{segment_distribution_single_gaussian_chain_approximated} in
Appendix \ref{equilibrium_distribution_functions_for_entangled_signle_gaussian_chain}).
Figure \ref{segment_number_distribution_function_single_chain_graph}
shows the segment number distribution functions for repulsive and
non-interacting slip-link models, for sufficiently large
$Z_{0}$. The distribution functions are normalized to
satisfy the following normalization condition.
\begin{equation}
 \int_{0}^{\infty} d(n/N_{0}) \, P_{\text{eq}}(n/N_{0}) = 1
\end{equation}
We can observe differences between two models clearly.
The difference is
especially large for small $n$. Eq
\eqref{segment_distribution_single_chain_definition_approximated_final}
is not a monotonically decreasing function and it has a maximum at $n /
N_{0} = 3 / 5$. Also,
eq \eqref{segment_distribution_single_chain_definition_approximated_final}
approaches to $0$ at the limit of $n \to 0$, while the distribution function
of the non-interacting slip-link model approaches to a non-zero constant.
Intuitively, this can be understood as the effect of the strong short range
repulsion between two neighboring slip-links.

\subsection{Bond Vector and Bond Length Distributions}
\label{bond_vector_and_bond_length_distributions}

\begin{figure}[c!]
 \centering
 {\includegraphics[width=0.95\linewidth,clip]{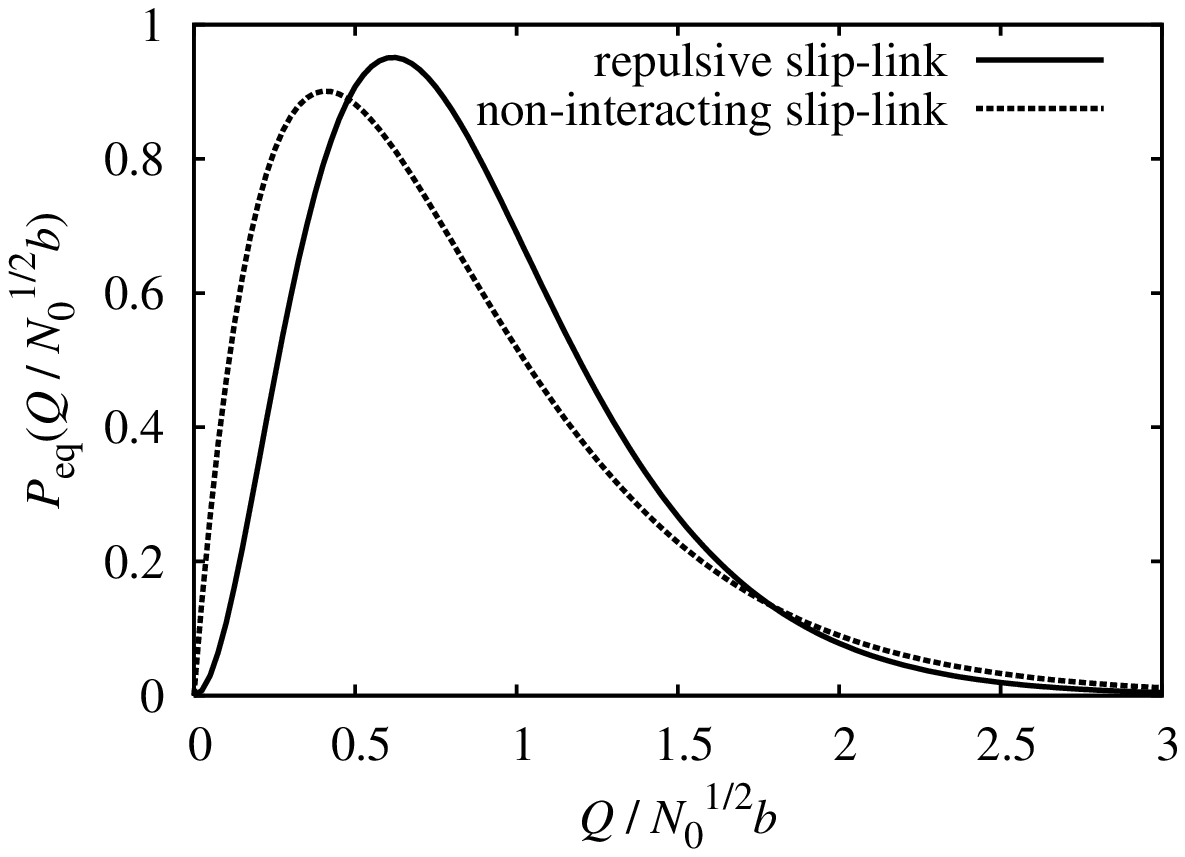}}
 \caption{}
 \label{bond_length_distribution_function_single_chain_graph}
\end{figure}

Finally we calculate the distribution functions for the bond vector and
bond length. Here we define the bond vector as the vector which
connects the two successive slip-links (or pair of a slip-link and a
chain end) on the same chain.
If we express the bond vector as $\bm{Q}$, the bond vector distribution
function $P_{\text{eq}}(\bm{Q})$ is expressed as follows.
\begin{equation}
 \label{bond_distribution_single_chain_definition}
  \begin{split}
   P_{\text{eq}}(\bm{Q}) & = \sum_{Z = 1}^{\infty}
   \int d\lbrace \bm{R}_{k} \rbrace d\lbrace N_{k} \rbrace \,
   \bigg[ \frac{1}{Z} \sum_{l = 1}^{Z} \delta(\bm{Q}
   - \bm{R}_{l} + \bm{R}_{l - 1}) \bigg]
   P_{\text{eq}}(\lbrace \bm{R}_{k} \rbrace,
   \lbrace N_{k} \rbrace, Z)   
  \end{split}
\end{equation}
Eq \eqref{bond_distribution_single_chain_definition} can be modified as
follows.
\begin{equation}
 \label{bond_distribution_single_chain_modified}
  \begin{split}
   P_{\text{eq}}(\bm{Q})
   & = \frac{\mathcal{V}}{\Xi \Lambda^{3}} \left(\frac{3}{2
   \pi b^{2}}\right)^{3 / 2}
   \frac{1}{N}
   \int_{0}^{N} dn \,
      \left(\frac{N}{n}\right)^{3/2}
   e^{  - 3 \bm{Q}^{2} / 2 n b^{2} } \\
   & \qquad \times  \sum_{Z = 1}^{\infty} \frac{\xi^{Z} }{N^{5 Z / 2 - 2}}
   \int d\lbrace N_{k} \rbrace \,
   \prod_{k = 1}^{Z} N_{k}^{3 / 2} \bigg[ \frac{1}{Z} \sum_{l = 1}^{Z} \delta(n - N_{l}) \bigg] \delta
   \bigg( N - \sum_{k = 1}^{Z} N_{k} \bigg) \\
  \end{split}
\end{equation}
The integrals over $\lbrace N_{k} \rbrace$ can be calculated in the same
way as the case of the segment number distribution function.
Then we have
\begin{equation}
 \label{bond_distribution_single_chain_modified2}
  \begin{split}
   P_{\text{eq}}(\bm{Q})
   & = \left(\frac{3}{2 \pi b^{2}}\right)^{3 / 2}
   \int_{0}^{N} dn \,
   \frac{1}{n^{3/2}}
   e^{  - 3 \bm{Q}^{2} / 2 n b^{2} } P_{\text{eq}}(n)
  \end{split}
\end{equation}
For sufficiently large $Z_{0}$,
the bond vector distribution function can be approximately expressed as
follows, by using eq \eqref{segment_distribution_single_chain_definition_approximated_final}.
\begin{equation}
 \label{bond_distribution_single_chain_approximated}
 \begin{split}
  P_{\text{eq}}(\bm{Q})
  & \approx \frac{25}{6} \sqrt{\frac{10}{\pi}} \left(\frac{3}{2 \pi N_{0} b^{2}}\right)^{3 / 2} \frac{1}{N_{0}}
  \int_{0}^{\infty} dn \,
   \exp \left( - \frac{3 \bm{Q}^{2}}{2 n b^{2}} - \frac{5 n}{2 N_{0}} \right) \\
  & = \frac{75}{2 \pi^{2}}
  \frac{|\bm{Q}|}{N_{0}^{2} b^{4}}
  K_{1} \bigg( \sqrt{\frac{15}{N_{0} b^{2}}} | \bm{Q}| \bigg) \\
 \end{split}
\end{equation}
where
$K_{1}(x)$ is the first order modified Bessel function of the second
kind \cite{Abramowitz-Stegun-book}.
(The detailed calculation is described in Appendix \ref{integrals_over_n}.)
From eq \eqref{bond_distribution_single_chain_approximated}, the bond
length distribution function (the bond length $Q$ is defined as $Q
\equiv |\bm{Q}|$) becomes
\begin{equation}
 \label{bond_length_distribution_single_chain_approximated}
  P_{\text{eq}}(Q)
  = 4 \pi Q^{2} P_{\text{eq}}(\bm{Q}) \approx \frac{150}{\pi}
  \frac{Q^{3}}{N_{0}^{2} b^{4}}
  K_{1} \left( \sqrt{\frac{15}{N_{0} b^{2}}} Q \right)
\end{equation}
From eq
\eqref{bond_length_distribution_single_chain_approximated} we
find that the form of the bond length distribution function is also
different from one of the single chain non-interacting slip-link model
(eq \eqref{bond_distribution_single_gaussian_chain_approximated} in
Appendix \ref{equilibrium_distribution_functions_for_entangled_signle_gaussian_chain}).
Figure \ref{bond_length_distribution_function_single_chain_graph}
shows the bond length distribution functions for repulsive and
non-interacting slip-link models, for sufficiently large
$Z_{0}$. The distribution functions are normalized to
satisfy the following normalization condition
\begin{equation}
 \int_{0}^{\infty} d(Q / \sqrt{N_{0} b^{2}}) \, P_{\text{eq}}(Q / \sqrt{N_{0} b^{2}}) = 1
\end{equation}
For small $Q$, the asymptotic form of eq
\eqref{bond_length_distribution_single_chain_approximated} becomes
\begin{equation}
 \label{bond_length_distribution_single_chain_asymptotic}
 P_{\text{eq}}(Q) \approx
 \frac{10 \sqrt{15}}{\pi} 
  \frac{Q^{2}}{N_{0}^{3/2} b^{3}} \propto Q^{2} \qquad (\text{for $Q \ll
  \sqrt{N_{0} b^{2}}$})
\end{equation}
This dependence of the asymptotic form on $Q$ is different from one for the
non-interacting slip-link model ($P_{\text{eq}}(Q) \propto Q$).
The difference between two
distributions is especially large for small $Q$. This trend is similar
to the difference between the segment number distribution functions.
The position of the maximum of eq
\eqref{bond_length_distribution_single_chain_approximated} is $Q / \sqrt{N_{0} b^{2}} \approx
0.616$ whereas it is $Q / \sqrt{N_{0} b^{2}} = 1 /
\sqrt{6} \approx 0.408$ for the non-interacting slip-link model. As
before, these properties can be understood as the effect of the
strong short range repulsion between slip-links.

\subsection{Comparison with PCN Simulations}

The distribution functions obtained in preceding subsections
(Sections \ref{sliplinkd_subchain_number_distribution}-\ref{bond_vector_and_bond_length_distributions}) are based
on the single subchain approximation. However, the validity of the single chain
approximation is not guaranteed for the PCN model.
In this subsection, we perform PCN simulations with
original and modified dynamic equations and directly calculate the
distribution functions.

The simulations are performed for linear monodisperse polymers with
$Z_{0} = 5, 10, 20$ and $40$. The osmotic terms to prevent aggregation of
slip-links\cite{Masubuchi-Ianniruberto-Greco-Marrucci-2008,Masubuchi-Uneyama-Watanabe-Ianniruberto-Greco-Marrucci-2010}
are added to
eqs \eqref{dynamic_equation_node_position}, \eqref{dynamic_equation_segment_number_flux},
and \eqref{dynamic_equation_segment_number_flux_modified}. The network
reconstruction is used to realize the steady state (that is the
equilibrium state for the modified model). The employed rule was NR1 in
Ref \citen{Masubuchi-Uneyama-Watanabe-Ianniruberto-Greco-Marrucci-2010}
(which is briefly explained in Section \ref{pcn_dynamic_equations}).
The unit cell dimension for periodic boundary condition is
$12 \sqrt{N_{0} b^{2}}$ and the total bond number in the cell is
ca. $10 \times 12^{3}$. The distribution functions are obtained for $10$
independent simulations and the averages are reported here for noise
reduction purposes.

\begin{figure}[c!]
 \centering
 {\includegraphics[width=0.95\linewidth,clip]{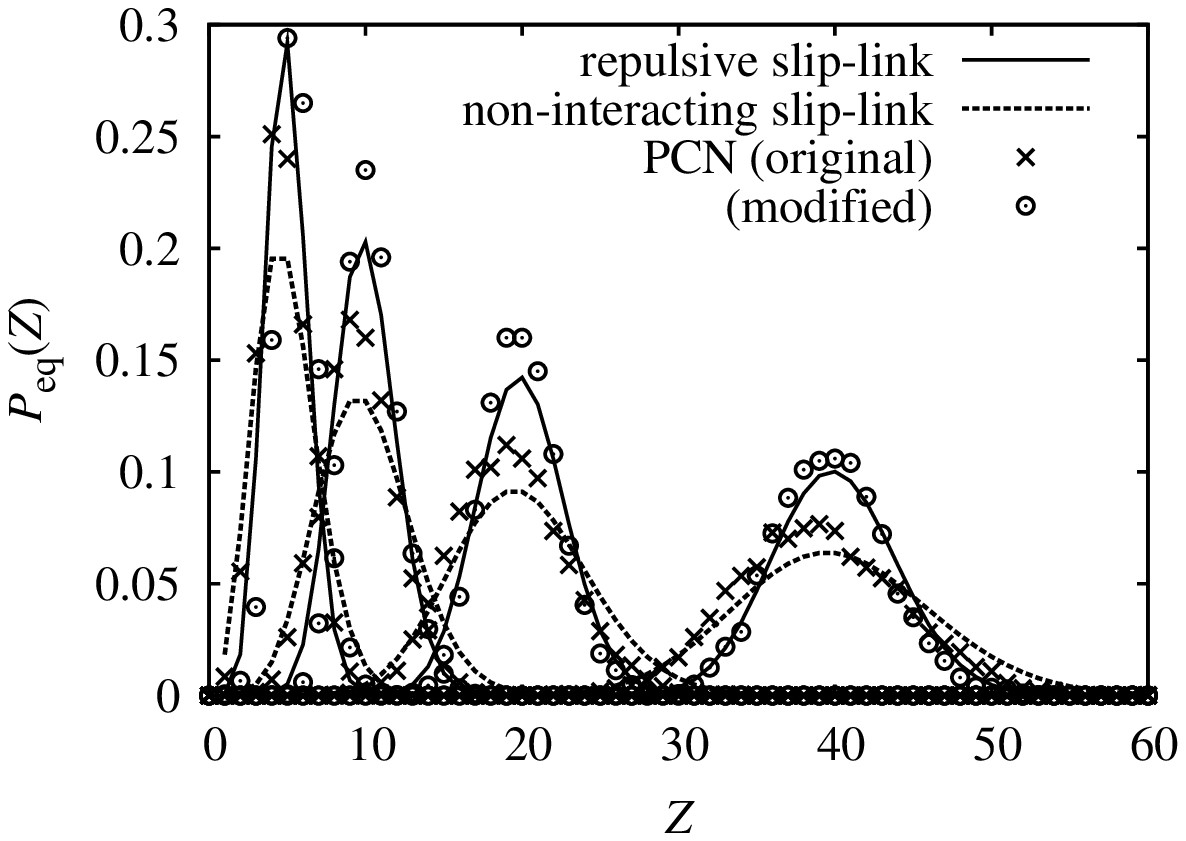}}
 \caption{}
 \label{sliplinked_number_distribution_function_comparison_graph}
\end{figure}

Figure \ref{sliplinked_number_distribution_function_comparison_graph}
shows the comparison of the subchain number distribution functions $P_{\text{eq}}(Z)$
calculated by the single chain models and the PCN simulations. As shown
in Figure
\ref{sliplinked_number_distribution_function_comparison_graph}, the
data of the modified PCN model agree well with the repulsive slip-link
model. The original PCN model gives a subchain number distribution
function which is broader than the repulsive slip-link model but still
sharper than the non-interacting slip-link model. Considering the
roughness of the single chain approximation, the agreement between
the modified PCN model and the repulsive slip-link model is rather surprising.
The CReTA primitive path extraction by Tzoumanekas and Theodorou
\cite{Tzoumanekas-Theodorou-2006}
also shows subchain number distributions which are
sharper than the Poissonian, but still broader than eq
\eqref{sliplinked_distribution_single_chain_approximated}. Thus these
models will be in between the non-interacting slip-link model
and the repulsive slip-link model (if we assume that the repulsive
interaction between slip-links is the main reason which cause
non-Poissonian distribution).

\begin{figure}[c!]
 \centering
 {\includegraphics[width=0.95\linewidth,clip]{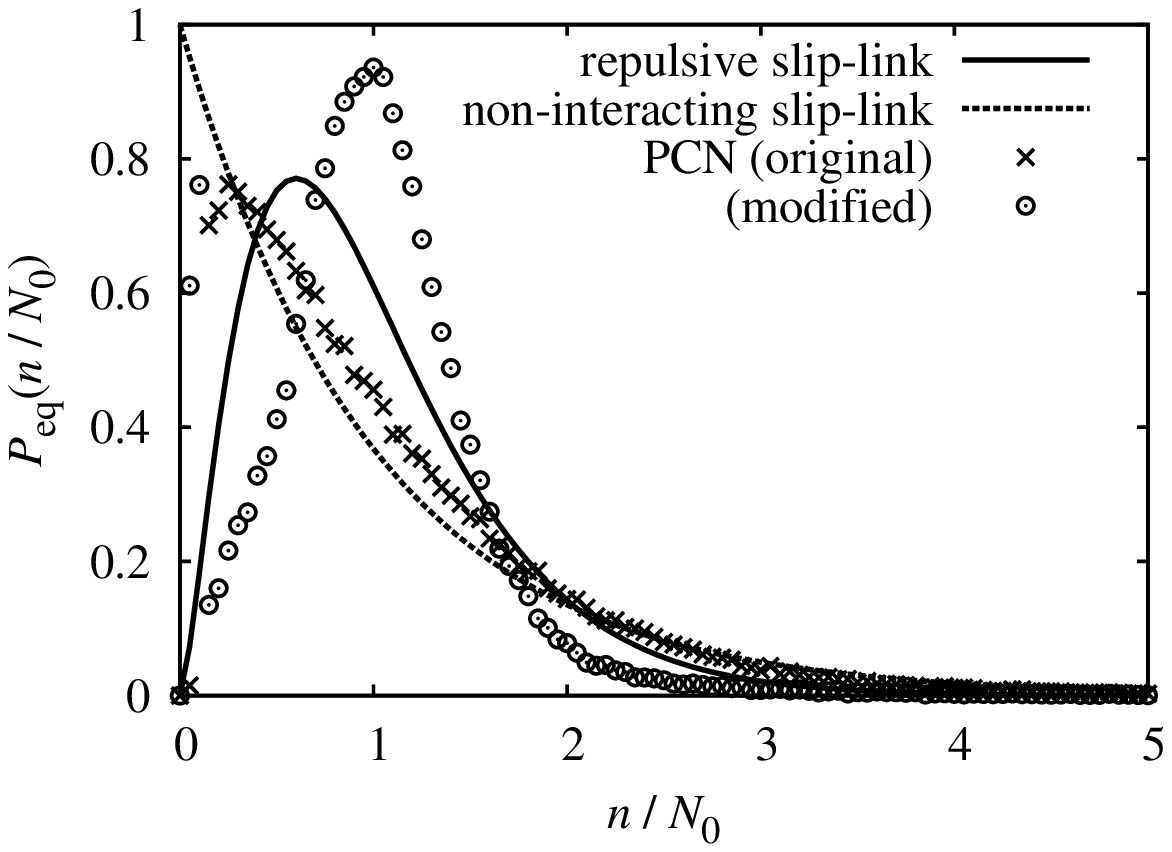}}
 \caption{}
 \label{segment_number_distribution_function_comparison_graph}
\end{figure}

Figure \ref{segment_number_distribution_function_comparison_graph} shows
the comparisons of the segment number distribution functions
$P_{\text{eq}}(n)$ for $Z_{0} = 40$. (The sharp peaks of $P_{\text{eq}}(n)$ by the PCN
simulations around $n / N_{0} = 0.1$ are the artifact due to the cutoff
of the segment number\cite{Masubuchi-Uneyama-Watanabe-Ianniruberto-Greco-Marrucci-2010}. In the followings we neglect these artificial
peaks.) Unlike the
case of the subchain number distribution, we can observe the deviation
of the modified PCN model data from the repulsive slip-link model. As
shown in Figure
\ref{segment_number_distribution_function_comparison_graph}, the segment
number distribution function of the modified PCN model is somehow shifted to
large $n$ region. However, the asymptotic behavior of the modified PCN
model at small $n$ is consistent
with the repulsive slip-link model ($P_{\text{eq}}(n) \propto n^{3/2}$).
The original PCN model gives the distribution between the
non-interacting and the repulsive slip-link models. This seems to be
qualitatively similar to the case of the subchain number distribution.
Here it is worth mentioning that the CReTA primitive path
extraction data \cite{Tzoumanekas-Theodorou-2006} gives the qualitatively
similar asymptotic behavior, $P(n) \to 0$  at $n \to 0$, to the modified
PCN model and the repulsive slip-link model. We will discuss the
segment number distribution functions later in detail (Section \ref{discussion_segment_number_distribution}).

\begin{figure}[c!]
 \centering
 {\includegraphics[width=0.95\linewidth,clip]{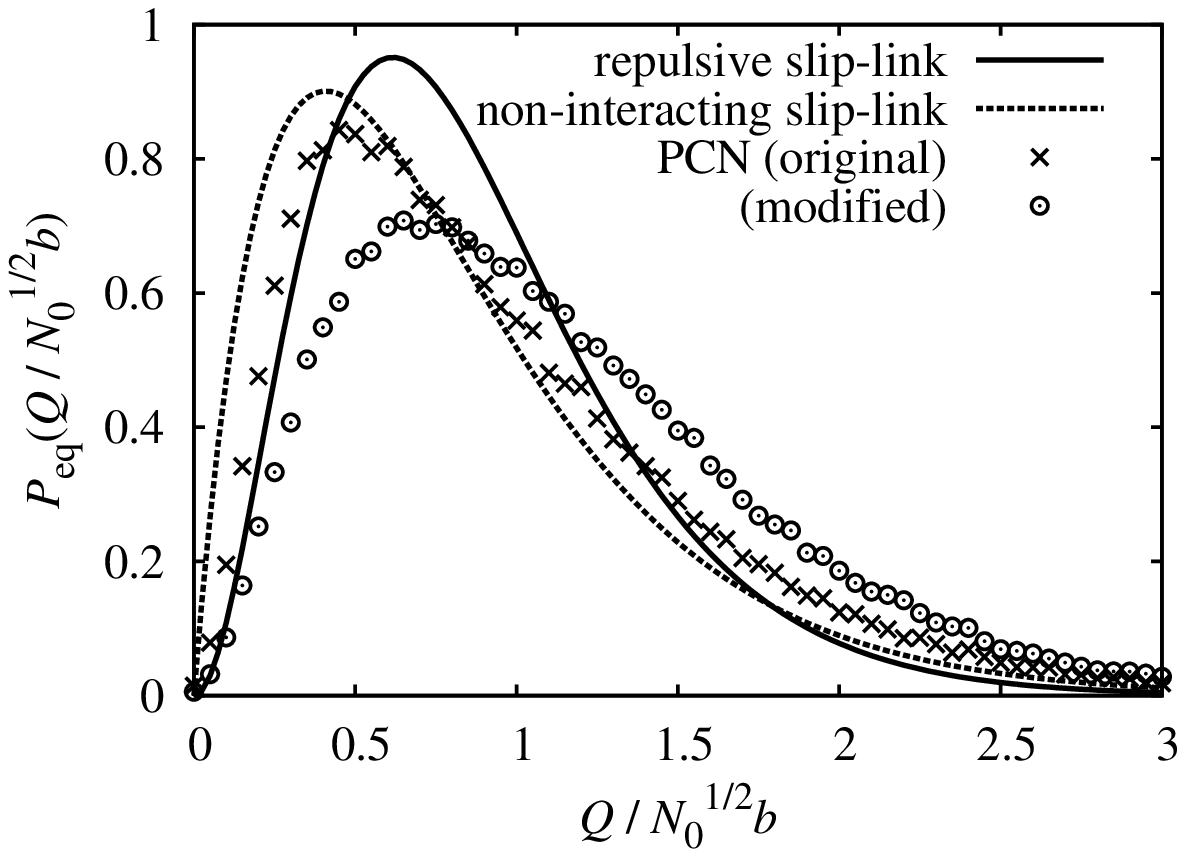}}
 \caption{}
 \label{bond_length_distribution_function_comparison_graph}
\end{figure}

Figure \ref{bond_length_distribution_function_comparison_graph} shows
the bond length distribution functions $P_{\text{eq}}(Q)$ for $Z_{0} = 40$.
We can find the bond length distribution function of the modified PCN
model deviates from the repulsive slip-link model. This result is physically
natural, because 
the bond length distribution function and the segment number
distribution function are expected to be related each other.
In the single chain model,
actually their relation is simply expressed by eq
\eqref{bond_distribution_single_chain_modified2}.
The bond length distribution function of the modified PCN model is slightly
shifted to large $Q$ region. The asymptotic behavior of the modified PCN
model at small $Q$ is also consistent with the repulsive slip-link model
($P_{\text{eq}}(Q) \propto Q^{2}$). The original PCN model data is again
between the non-interacting and repulsive slip-link models.

We can observe that both the original and modified PCN simulation data
have relatively large probability for large $Q$.
This implies that subchains in the system are
stretched compared with the ideal Gaussian subchains. This would
be because chains are required to form a connected, tetra-functional
network structure in the PCN model. (In the PCN model, a slip-link
spatially binds two chains, and the bound chains will be somehow
stretched.) This can be interpreted as a result of the {\em interchain 
force-balance} condition imposed in the PCN model (this condition is
automatically imposed by eqs
\eqref{connectivity_information_node_position} and
\eqref{dynamic_equation_node_position}).
The possible chain conformations are
limited compared with the single chain model, where no such constraint
exists.
As a result, the bond length distribution function becomes
broader and deviates from the distribution function in the single chain
model. This is qualitatively consistent with the simulation result.
Also, the segment number distribution would shift to large
segment number direction, which is again qualitatively consistent with
the simulation result.
Therefore, we expect the constraint for chains to be the origin
of the deviation of $P_{\text{eq}}(Q)$.
The origin of the deviation of $P_{\text{eq}}(n)$ will be also the
same.

In the single chain model, effects of the surrounding chains are
expressed only by the effective chemical potential $\epsilon$.
Such a rough approximation would not be fully applicable to describe the
statistics of slip-linked multi chains. The effective chemical potential
controls only the number of slip-linked subchains, and the
statistics of bond vectors or segment numbers are determined only by the
effective free energy. If interactions between the target chain and the
surrounding chains are not simple and cannot be described by a
single parameter $\epsilon$, the single chain model cannot reproduce the
PCN simulation results.
To analyze the statistical properties of the PCN
model theoretically, therefore, we will need to improve the single chain
approximation to take account the effects of surrounding chains.
For example, some sort of spatial correlation effect may be took into
account to the model, by using the connectivity information (such as 
the connectivity map $C$ in the PCN model).
Several radial distribution functions of nodes are obtained by the CReTA
primitive path extraction\cite{Tzoumanekas-Theodorou-2006}.
Kindt and Briels\cite{Kindt-Briels-2007} modeled that the
average number of entanglements between two chains as a function of the
distance between centers of mass of chains for their highly
coarse-grained single particle model of entangled polymers. If we use of the connectivity
information and reproduce these spatial correlations, several
statistical properties may be improved.
Of course, there are other possible factors which affect the statistical
properties of the PCN model. 
(As we mentioned, the network reconstruction
rule\cite{Masubuchi-Uneyama-Watanabe-Ianniruberto-Greco-Marrucci-2010}
affects several statistical properties of the PCN model.
Recent simulation results show that the spatial correlation between
nodes is somewhat affected by the form of the interaction potential
between nodes, although rheological properties are almost
unchanged\cite{Okuda-Inoue-Masubuchi-Uneyama-Hojo-submitted}. Such a
local structural change may affect the distribution functions to some
extent. If we directly control the spatial correlations of slip-links by
controlling the reconstruction rules and the connectivity information,
the distribution functions will be affected.)

\section{Discussion}
\label{discussion}

\subsection{Detailed Balance Condition and Effective Free Energy}

As shown in Section
\ref{langevin_form_of_segment_transport_equation_and_effective_free_energy},
the PCN dynamic equations can be reduced to the detailed-balanced Langevin
equations with two modifications.
It is rather surprising that simple modifications can reproduce
rather complicated, detailed-balanced Langevin equations.
The modified (detailed-balanced) version of the PCN
model can reproduce several probability distribution functions which are
qualitatively similar to the ones predicted by the single chain repulsive slip-link model.

Here we consider the original PCN dynamic equation
for $\lbrace N_{i,k} \rbrace$ (eqs
\eqref{dynamic_equation_segment_number} and \eqref{dynamic_equation_segment_number_flux}) again.
The fact that we only need relatively minor modifications for the
original PCN dynamics equations implies that the original PCN dynamics
itself is already a reasonable approximation for a
statistical mechanically sound model.
Actually, we can show that the original PCN model approximately satisfies the detailed balance
condition under a certain condition. We
consider the most probable value of $N_{i,k}$
under a given $\lbrace \bm{R}_{i,k} \rbrace$ and $Z_{i}$. The most
probable state $\lbrace \bar{N}_{i,k} \rbrace$ minimizes the effective
free energy \eqref{effective_free_energy_pcn}. $\bar{N}_{i,k}$ is simply
given as
\begin{equation}
 \bar{N}_{i,k} = \bar{\rho}_{i} |\bm{R}_{i,k} - \bm{R}_{i,k - 1}|
\end{equation}
where $\bar{\rho}_{i}$ is the most probable value of the monomer density
along the chain on
the $i$-th chain ($\bar{\rho}_{i,k}$ is independent of $k$ and thus
we simply describe it as $\bar{\rho}_{i}$).
If the deviation of $N_{i,k}$ from the most probable value is small, we
can rewrite $N_{i,k}$ as $N_{i,k} = \bar{N}_{i,k} + \delta N_{i,k}$ and
expand the free energy or the dynamic equation into power series of $\lbrace
\delta N_{i,k} \rbrace$. Expanding eqs
\eqref{dynamic_equation_segment_number_flux_alternative},
\eqref{line_density_definition_alternative} and
\eqref{effective_free_energy_pcn} into power
series and retaining only the leading order terms, we have
\begin{equation}
 \label{effective_free_energy_pcn_expanded}
  \mathcal{F}_{\text{eff}}(\lbrace \bm{R}_{i,k} \rbrace, \lbrace
  N_{i,k} \rbrace, \lbrace Z_{i} \rbrace)
   \approx \frac{3 k_{B} T} {2 b^{2}} \sum_{i} \sum_{k = 1}^{Z_{i}}
   \bigg[ 
   \frac{(\bm{R}_{i,k} -
  \bm{R}_{i,k - 1})^{2}}{\bar{N}_{i,k}} + \frac{\delta
  N_{i,k}^{2}}{\bar{\rho}_{i}^{2} \bar{N}_{i,k}} \bigg]
\end{equation}
\begin{equation}
 \label{line_density_definition_alternative_expanded}
  \rho_{i,k}(\lbrace \bm{R}_{i,k} \rbrace, \lbrace N_{i,k} \rbrace)
  \approx \bar{\rho}_{i}
\end{equation}
\begin{equation}
 \label{dynamic_equation_segment_number_flux_alternative_expanded}
   J_{i,k}(t)
   \approx
    \frac{2 \bar{\rho}_{i}^{2}}{\zeta} \frac{3 k_{B} T}{b^{2}}
   \left[ 
   \frac{\delta N_{i,k}}{\bar{\rho}_{i}^{2} \bar{N}_{i,k}}
    - \frac{\delta N_{i,k + 1}}{\bar{\rho}_{i}^{2} \bar{N}_{i,k + 1}} \right]
   + \sqrt{\frac{4 \bar{\rho}^{2}_{i} k_{B} T}{\zeta}} w^{(N)}_{i,k}(t)
\end{equation}
The original PCN model, eqs  \eqref{line_density_definition} and
\eqref{dynamic_equation_segment_number_flux} also reduce to eqs
\eqref{line_density_definition_alternative_expanded} and
\eqref{dynamic_equation_segment_number_flux_alternative_expanded},
respectively.
Therefore, the original PCN model satisfies the
detailed balance condition if $\lbrace N_{i,k} \rbrace$ is {\em in the
vicinity of the most probable state} $\lbrace \bar{N}_{i,k} \rbrace$.
(Intuitively, this is because the segment density $\rho_{i,k}$ is not sensitive to the
node index $k$ in the vicinity of the most
probable state. The harmonic and arithmetic averages are almost the same
and the spurious drift velocity is negligibly small, under this condition.)
Unfortunately, this condition is
not always satisfied even in equilibrium, because $\delta N_{i,k}$ is generally
not negligibly small due to the thermal fluctuation.
Nevertheless, the probability that $\lbrace N_{i,k} \rbrace$ is close to
$\lbrace \bar{N}_{i,k} \rbrace$ is expected not to be small. Thus we
consider that the analysis performed here qualitatively explains why
the original PCN model works as a good approximation.

The new definition for the segment density $\rho_{i,k}$
(eq \eqref{line_density_definition_alternative}) proposed in this work
is expressed as the {\em harmonic average} of the local densities in two neighboring
bonds. Although this form is not intuitive, from the view point of the
discretization scheme, it is not unreasonable.
The PCN dynamic equation for $\lbrace N_{i,k} \rbrace$ (eqs \eqref{dynamic_equation_segment_number} and
\eqref{dynamic_equation_segment_number_flux}) is interpreted as the
stochastic diffusion equation on one dimensional discrete lattice.
At the continuum limit, the PCN dynamic equation will reduce to
the stochastic partial differential equation\cite{Dean-1996} which contains the
multiplicative noise. To discretize the multiplicative
noises appear in the stochastic partial differential equations,
non-arithmetic averages are sometimes preferred than the simple
arithmetic average.
For example, the multiplicative noise in the dynamic density functional
equation can be accurately discretized with the geometric average
form\cite{Uneyama-2007}. (The geometric or harmonic average is zero
if one of the averaged numbers is zero, whereas the arithmetic average is
nonzero. Such a property is required to generate the multiplicative
noise accurately and stably.)
Thus we expect that the harmonic average form is more natural than the
arithmetic average form.

We should note that the statistical properties of the PCN network
reconstruction rules
\cite{Yaoita-Isaki-Masubuchi-Watanabe-Ianniruberto-Greco-Marrucci-2008,Masubuchi-Uneyama-Watanabe-Ianniruberto-Greco-Marrucci-2010}
are not considered in this work. The network reconstruction rules
proposed so far
do not satisfy the detailed balance condition and therefore the PCN
model is not fully detailed-balanced, even if we employ the modified
dynamic equations.
(Especially for small $Z_{0}$, the network reconstruction
rule affects the statistical properties significantly.)
To understand the statistical properties of the full PCN dynamics, the
statistical mechanical analysis of the network reconstruction rules is
also required. It seems to be difficult to satisfy the detailed
balance condition in the network reconstruction process of multi chain
slip-link models due to its intrinsic complicatedness, and thus it is left for a
future work. (The exact equilibrium
statistics of a slip-linked multi chain system is not simple nor
intuitive \cite{Uneyama-Horio-2011}, unlike a single chain system.)
Although we do not consider further
modifications here, the current work
will be useful to find strategies for other modifications which
improve the PCN model.

\subsection{Segment Number Distribution}
\label{discussion_segment_number_distribution}

As we showed and discussed in previous sections, the effective repulsive
interaction between slip-links (eq
\eqref{effective_potential_pcn_sliplink}) affects statistical
properties of the network qualitatively.
We expect that the statistics can be further changed by changing the
interaction potential (or interaction strength) between slip-links.
In fact, if we consider the limit of the strong repulsion,
distribution functions become much different from non-interacting
slip-link model (see Appendix \ref{equilibrium_distribution_functions_at_strong_repulsion_limit}).
Roughly speaking, the variance of the subchain number distribution
function can be used to estimate the strength the effective repulsive
interaction between slip-links. It is interesting to compare our model
(or our picture) with other models or primitive path extraction results,
and discuss whether our model is similar to others or not.


Among the examined distribution functions, the segment number
distribution function seems to be the most sensitive to the interaction
between slip-links.
The segment number distribution of the single chain repulsive slip-link
model is qualitatively much different
from one of the single chain non-interacting slip-link model.
The segment number distribution functions which have similar forms to eq
\eqref{segment_distribution_single_chain_definition_approximated_final}
are already
proposed by Tzoumanekas and Theodorou\cite{Tzoumanekas-Theodorou-2006},
and Greco\cite{Greco-2008}.

Here we compare our model with these
distribution functions.
Tzoumanekas and Theodorou proposed the following empirical form for the
segment number distribution, to fit their CReTA primitive path
extraction data.
\begin{equation}
 \label{segment_distribution_single_chain_definition_tzoumanekas_theodorou}
 P_{\text{eq}}(n) = \frac{1}{N_{0}}\frac{\tilde{b} \tilde{c}}{\tilde{c} - \tilde{b}}
 ( e^{- \tilde{b} n / N_{0}} - e^{- \tilde{c} n / N_{0}} )
\end{equation}
Here $1 < \tilde{b} < 2$ is a fitting parameter and $\tilde{c} = \tilde{b} /
(\tilde{b} - 1)$. They reported that the segment number distribution
functions obtained by the CReTA can be fitted well to eq
\eqref{segment_distribution_single_chain_definition_tzoumanekas_theodorou}
with $\tilde{b} = 1.30$.
Eq \eqref{segment_distribution_single_chain_definition_tzoumanekas_theodorou}
can be also utilized to fit the distribution functions by the original
PCN model with the fitting parameter $\tilde{b} = 1.96$
\cite{Masubuchi-Uneyama-Watanabe-Ianniruberto-Greco-Marrucci-2010}. This
value is close to $2$, and for such a case eq
\eqref{segment_distribution_single_chain_definition_tzoumanekas_theodorou}
can be approximated well by the following simple form.
\begin{equation}
 \label{segment_distribution_single_chain_definition_tzoumanekas_theodorou_limit}
 P_{\text{eq}}(n) \to \frac{4 n}{N_{0}^{2}} e^{- 2 n / N_{0}} \qquad
 (\tilde{b} \to 2)
\end{equation}
Eq
\eqref{segment_distribution_single_chain_definition_tzoumanekas_theodorou_limit} is
quite similar to eq
\eqref{segment_distribution_single_chain_definition_approximated_final}.
Tzoumanekas and Theodorou proposed that the physical origin of such a
non-monotonic form of eq 
\eqref{segment_distribution_single_chain_definition_tzoumanekas_theodorou}
is the effective repulsion (or the blocking effect) between topological
constraints, which depends the chemical distance.
This is qualitatively the same as our repulsive slip-link model.
Of course, the CReTA primitive path extraction results and their
molecular model are different from the PCN model and our slip-link model.
For example, the CReTA primitive path data are taken
for an atomistic polymer model while the PCN model employs a rather
coarse-grained Gaussian chain model. Besides, it is not clear whether
the slip-links in the PCN model and the topological constraints extracted by the CReTA
have the same physical properties.
Thus the distribution function by the PCN model does not necessarily to be
coincide to one by the CReTA.
Actually, at least the values of the
fitting parameter $\tilde{b}$ for two distributions are different.
Nonetheless we consider that the similarity between the Tzoumanekas-Theodorou
model and our repulsive slip-link model implies that there are some common
properties of slip-links (or topological constraints).
(We expect that the detailed form of
the effective repulsive potential between topological constraints in the
Tzoumanekas-Theodorou model is different from our repulsive slip-link
model. Judging from the distribution functions, the repulsive
interaction in their model seems to be weaker than our model. But the
detail repulsion mechanisms in the Tzoumanekas-Theodorou model or the
CReTA results are not so clear. Further investigations will be future works.)

On the other hand, in the Greco model, the segment number
distribution function is given as
\begin{equation}
 \label{segment_distribution_single_chain_definition_greco}
  P_{\text{eq}}(n) \propto s(\tilde{m}) \sqrt{\frac{n}{N_{0}}} e^{- \tilde{m}^{2} n
  / N_{0}} + \sqrt{\pi}
  \left[ \frac{1}{2} + s^{2}(\tilde{m}) \frac{n}{N_{0}} \right]
   e^{(1 / \tilde{m}^{2} - 2) n / N_{0}} \erfc\left(- s(\tilde{m}) \sqrt{\frac{n}{N_{0}}}\right)
\end{equation}
where $\tilde{m} \ge 1$ is the renormalized chemical potential (which is used as a fitting parameter), $s(x) \equiv (x^{2} -
1) / x$, and $\erfc(x)$ is the complementary error
function\cite{Abramowitz-Stegun-book}.
Eq \eqref{segment_distribution_single_chain_definition_greco} has
a similar form to eq
\eqref{segment_distribution_single_chain_definition_tzoumanekas_theodorou}\cite{Greco-2008,Masubuchi-Uneyama-Watanabe-Ianniruberto-Greco-Marrucci-2010}.
In the Greco model, the slip-links are non-interacting.
The non-monotonic nature of eq
\eqref{segment_distribution_single_chain_definition_greco} comes from
the fluctuation effect due to the smallness of the system.
This is qualitatively different from our model.
It is interesting that different mechanisms (the fluctuation effect and
the repulsive interaction between slip-links) give the similar result.
Further progress of theories would be required to understand the origin
of non-monotonic segment number distribution functions.

Although there will be other factors (the osmotic
force between nodes, the network reconstruction, or
constraint to form the tetra-functional network structure), the
interaction between slip-links is one important factor which determines
the network statistics.
The concept of the effective repulsive interaction between slip-links
will be also useful to consider statistical properties of other models
or methods such as the primitive path extraction methods
\cite{Everaers-Sukumaran-Grest-Svaneborg-Sivasubramanian-Kremer-2004,Kroger-2005,Tzoumanekas-Theodorou-2006}
or other slip-link based models without explicit free energy model \cite{Doi-Takimoto-2003}.
For example, the differences among various primitive
path extraction methods may be related to the differences of the
effective slip-link interaction potential models.
Once we obtain the explicit form of the effective interaction potential,
it will be possible to tune it so that the model reproduces the required
equilibrium statistics.
We can employ the effective interaction potential for
slip-links determined
from the network statistics by the primitive path
extraction methods to make the slip-link models compatible with the
primitive path extraction data.
(We can
easily replace the effective free energy model by other models to
tune the statistics of the modified PCN model, because the forces in the
modified PCN model are expressed as variational forms.)

\section{Conclusions}
\label{conclusion}

We showed that the PCN dynamic equations do not satisfy the
detailed-balance condition and therefore the PCN model does not have the
thermal equilibrium state.
However, by introducing heuristic modifications, the PCN dynamics can be
recovered to satisfy the detailed
balance condition. We proposed two modifications. One is to change the
definition of the segment density along the chain to the harmonic
average form (eq \eqref{line_density_definition_alternative}), and
another is to add the spurious drift term to the monomer flux equation
(eq \eqref{dynamic_equation_segment_number_flux_alternative}).
From the modified PCN model, we obtained the effective free energy from
which the equilibrium statistical properties of the PCN model are
determined. The effective PCN free energy has a different form from
the free energy of ideal Gaussian chains. This can be understood that
the effective PCN free energy contains the contribution from the repulsive
interaction between slip-links.

To analyze the equilibrium statistical properties of the PCN model, we
constructed a single chain model with repulsive slip-links.
The equilibrium distribution functions derived from the single
chain repulsive slip-link model are qualitatively different from ones for the single chain
non-interacting slip-link model, due to the repulsive interaction between slip-links.
It was shown that the slip-linked subchain number distribution function for the
repulsive slip-link model is much sharper than one for non-interacting slip-links.
The interaction between slip-links strongly affects the segment number
distribution function, especially if the segment number is small.
The equilibrium properties obtained from the single chain slip-link
model are qualitatively similar to the PCN simulation results, although
the agreement is not perfect. We consider the repulsive interaction
between slip-links is an important factor which determines the
statistical properties of the PCN model.
The repulsive slip-link picture would be useful to
improve the model statistics or design a new model.

\section*{Acknowledgment}

This work is supported by JST-CREST.

\section*{Appendix}
\appendix

\section{Single Chain Non-Interacting Slip-Link Model}
\label{equilibrium_distribution_functions_for_entangled_signle_gaussian_chain}

In this appendix, we show a brief derivation of the equilibrium
distribution functions for the single chain non-interacting slip-link model
\cite{Schieber-2003}. (The derivation mainly follows Ref
\citen{Schieber-2003} but the expressions are slightly modified to allow
the direct comparison with the repulsive slip-link model in the main text.)
The free energy of a Gaussian chain with non-interacting slip-links is
expressed as the single chain version of eq \eqref{free_energy_gaussian_chains}.
\begin{equation}
 \label{free_energy_single_gaussian_chain}
  \mathcal{F}_{0}(\lbrace \bm{R}_{k} \rbrace, \lbrace N_{k} \rbrace, Z)
  \equiv \frac{3 k_{B} T} {2 b^{2}} \sum_{k = 1}^{Z} \frac{(\bm{R}_{k} -
  \bm{R}_{k - 1})^{2}}{N_{k}}
  + \frac{3 k_{B} T}{2} \sum_{k = 1}^{Z} \ln N_{k}
\end{equation}
As we mentioned in the main text, the Gaussian free energy
contains a term which is proportional to $\ln N_{i}$ while
the PCN effective free energy \eqref{effective_free_energy_pcn}
does not.

The grand potential is defined as follows, by introducing the
effective chemical potential for a subchain, $\epsilon_{0}$.
\begin{equation}
 \label{grand_potential_single_gaussian_chain}
  \mathcal{J}_{0}(\lbrace \bm{R}_{k} \rbrace, \lbrace N_{k} \rbrace, Z)
   \equiv \mathcal{F}_{0}(\lbrace \bm{R}_{k} \rbrace, \lbrace N_{k}
   \rbrace, Z) - \epsilon_{0} Z \\
\end{equation}
The equilibrium statistical probability can be written as
\begin{equation}
 P_{\text{eq}}(\lbrace \bm{R}_{k} \rbrace, \lbrace N_{k} \rbrace, Z) = \frac{1}{\Xi_{0}
  \Lambda^{3 (Z + 1)} N^{Z - 1}}
 \delta \bigg( N - \sum_{k = 1}^{Z} N_{k} \bigg)
  \exp \left[ - \frac{\mathcal{J}(\lbrace \bm{R}_{k} \rbrace,
	\lbrace N_{k} \rbrace, Z)}{k_{B} T} \right]
\end{equation}
with $\Xi_{0}$ being the grand partition function. The grand partition
function reduces to a simple form.
\begin{equation}
 \label{partition_function_single_gaussian_chain}
\begin{split}
 \Xi_{0}
 & \equiv \sum_{Z = 1}^{\infty} \frac{1}{\Lambda^{3 (Z + 1)} N^{Z - 1}}
 \int d\lbrace \bm{R}_{k} \rbrace d\lbrace N_{k} \rbrace \,
 \delta \bigg( N - \sum_{k = 1}^{Z} N_{k} \bigg)
 \exp \left[ - \frac{\mathcal{J}(\lbrace \bm{R}_{k} \rbrace,
       \lbrace N_{k} \rbrace, Z)}{k_{B} T} \right] \\
 & = \frac{\mathcal{V}}{\Lambda^{3}} \xi_{0} e^{\xi_{0}}
\end{split}
\end{equation}
Here we defined the dimensionless effective fugacity $\xi_{0}$ as
\begin{equation}
 \xi_{0} \equiv e^{\epsilon_{0} / k_{B} T}
 \left(\frac{2 \pi b^{2}}{3 \Lambda^{2}}\right)^{3 / 2}
\end{equation}
The slip-linked subchain number distribution function becomes the Poisson distribution.
\begin{equation}
 \label{sliplinked_distribution_single_gaussian_chain}
  P_{\text{eq}}(Z)
  = \int d \lbrace \bm{R}_{k} \rbrace d\lbrace N_{k} \rbrace \,
 P_{\text{eq}}(\lbrace \bm{R}_{k} \rbrace,
       \lbrace N_{k} \rbrace, Z)
  = \frac{1}{(Z - 1)!} \xi_{0}^{Z - 1} e^{-\xi_{0}}
\end{equation}
Eq \eqref{sliplinked_distribution_single_gaussian_chain} can be
rewritten by using the relation, $\langle Z \rangle_{\text{eq}} = Z_{0} = \xi_{0} + 1$.
\begin{equation}
 \label{sliplinked_distribution_single_gaussian_chain_modified}
  P_{\text{eq}}(Z)
  = \frac{1}{(Z - 1)!} (Z_{0} - 1)^{Z - 1} e^{- (Z_{0} - 1)}
\end{equation}
The variance of $Z$
is related to the average of $Z$ as
\begin{equation}
 \label{variance_z_single_gaussian_chain}
  \langle (Z - Z_{0})^{2} \rangle_{\text{eq}}
  = Z_{0} - 1 \approx Z_{0} \qquad (\text{for $Z_{0} \gg 1$})
\end{equation}

The segment number distribution function is expressed as
\begin{equation}
 \label{segment_distribution_single_gaussian_chain}
 \begin{split}
  P_{\text{eq}}(n)
  & = \sum_{Z = 1}^{\infty}
\int d \lbrace \bm{R}_{k} \rbrace d\lbrace N_{k} \rbrace \,
  \bigg[ \frac{1}{Z} \sum_{l = 1}^{Z} \delta(n - N_{l}) \bigg]
  P_{\text{eq}}(\lbrace \bm{R}_{k} \rbrace,
       \lbrace N_{k} \rbrace, Z) \\
  & = e^{-\xi_{0}} \sum_{Z = 1}^{\infty} \frac{\xi_{0}^{Z - 1}}{Z N} \sum_{l = 1}^{Z} \frac{1}{N^{Z - 2}}
  \int d\lbrace N_{k} \rbrace \,
  \delta(n - N_{l}) \delta \bigg( N - \sum_{k = 1}^{Z} N_{k} \bigg) \\
 \end{split}
\end{equation}
The integrals over $\lbrace N_{k} \rbrace$ can be calculated to be
\begin{equation}
  \frac{1}{N^{Z - 2}} \int d\lbrace N_{k} \rbrace \,
  \delta(n - N_{l}) \delta \bigg( N - \sum_{k = 1}^{Z} N_{k} \bigg)  =
  \begin{cases}
   \displaystyle N \delta(n - N) & (Z = 1) \\
   \displaystyle \frac{1}{(Z - 2)!} \left(1 - \frac{n}{N}\right)^{Z - 2} & (Z \ge 2)
  \end{cases}
\end{equation}
Thus eq \eqref{segment_distribution_single_gaussian_chain} can be
reduced to the following form.
\begin{equation}
 \label{segment_distribution_single_gaussian_chain_modified}
  \begin{split}
   P_{\text{eq}}(n)
   & = e^{- (Z_{0} - 1)} \delta\left(n - N \right)
   + \frac{Z_{0} - 1}{N} \exp \left[ - (Z_{0} - 1) \frac{n}{N} \right] \\
   & = e^{1 - N / N_{0}} \delta(n - N)
   + \frac{N - N_{0}}{N_{0} N} e^{- n (N - N_{0}) / N N_{0}}
  \end{split}
\end{equation}
where we used $Z_{0} = N / N_{0}$ (notice that, this $N_{0}$
is different from $N_{e}$ in Ref
\citen{Schieber-2003}).
For large $Z_{0}$, eq
\eqref{segment_distribution_single_gaussian_chain_modified} can be
approximated by the exponential function.
\begin{equation}
 \label{segment_distribution_single_gaussian_chain_approximated}
  P_{\text{eq}}(n) \approx \frac{1}{N_{0}} e^{-n / N_{0}}
\end{equation}

The bond vector distribution function can be calculated in a similar way.
\begin{equation}
 \label{bond_distribution_single_gaussian_chain}
 \begin{split}
  P_{\text{eq}}(\bm{Q})
  & = \sum_{Z = 1}^{\infty} \int d\lbrace \bm{R}_{k} \rbrace
  d\lbrace N_{k} \rbrace \,
  \bigg[ \frac{1}{Z} \sum_{l = 1}^{Z} \delta(\bm{Q} - \bm{R}_{l}
  + \bm{R}_{l - 1}) \bigg] P_{\text{eq}}(\lbrace \bm{R}_{k} \rbrace,
       \lbrace N_{k} \rbrace, Z) \\
  & = \left(\frac{3}{2 \pi b^{2}}\right)^{3/2} \int_{0}^{N} dn \,
  \frac{1}{n^{3/2}} e^{- 3 \bm{Q}^{2} / 2 n b^{2}} P_{\text{eq}}(n) \\
 \end{split}
\end{equation}
Substituting eq
\eqref{segment_distribution_single_gaussian_chain_modified} into eq 
\eqref{bond_distribution_single_gaussian_chain}, we have
\begin{equation}
 \label{bond_distribution_single_gaussian_chain_modified}
 \begin{split}
  P_{\text{eq}}(\bm{Q})
  & = \left(\frac{3}{2 \pi N b^{2}}\right)^{3/2} e^{- 3
  \bm{Q}^{2} / 2 N b^{2} - N / N_{0} + 1} \\
  & \qquad + \left(\frac{3}{2 \pi N_{0} b^{2}}\right)^{3/2}
  \frac{N - N_{0}}{N_{0} N}
  \int_{0}^{N} dn \,
  \left(\frac{N_{0}}{n}\right)^{3/2}
  \exp \left[ - \frac{3 \bm{Q}^{2}}{2 n b^{2}} - 
  \frac{N - N_{0}}{N_{0} N} n \right] \\
  & = \left(\frac{3}{2 \pi N b^{2}}\right)^{3/2} e^{- 3
  \bm{Q}^{2} / 2 N b^{2} - N / N_{0} + 1}
  + \frac{3}{4 \pi b^{2}}
  \frac{N - N_{0}}{N_{0} N}
   \frac{1}{|\bm{Q}|} \\
  & \qquad \times 
  \bigg[ \exp\bigg(\sqrt{\frac{6}{b^{2}}
  \frac{N - N_{0}}{N_{0} N}} |\bm{Q}| \bigg) \erfc\bigg(\sqrt{\frac{3
  }{2 N b^{2}}} |\bm{Q}| + \sqrt{\frac{N}{N_{0}} -
   1}\bigg) \\
  & \qquad \qquad + \exp\bigg(- \sqrt{\frac{6}{b^{2}}
  \frac{N - N_{0}}{N_{0} N}} |\bm{Q}| \bigg)
  \erfc\bigg(\sqrt{\frac{3}{2 N b^{2}}} |\bm{Q}| - \sqrt{\frac{N}{N_{0}} -
   1}\bigg) \bigg]
 \end{split}
\end{equation}
where $\erfc x$ is the complementary error function \cite{Abramowitz-Stegun-book}.
(See Appendix \ref{integrals_over_n} for details.)
Although eq \eqref{bond_distribution_single_gaussian_chain_modified}
is exact, it is quite complicated and not intuitive. It can be
reduced to the following
simple approximate form for sufficiently large $Z_{0}$.
\begin{equation}
 \label{bond_distribution_single_gaussian_chain_approximated}
 P_{\text{eq}}(\bm{Q})
 \approx
 \frac{3}{2 \pi N_{0} b^{2} |\bm{Q}|} \exp \bigg( -
 \sqrt{\frac{6}{N_{0} b^{2}}} |\bm{Q}| \bigg)
\end{equation}
The bond length distribution function can be expressed as
\begin{equation}
 \label{bond_length_distribution_single_gaussian_chain_approximated}
 P_{\text{eq}}(Q) 
 \approx
 \frac{6 Q}{N_{0} b^{2}}
 \exp \bigg( - \sqrt{\frac{6}{N_{0} b^{2}}} Q \bigg)
\end{equation}
Eq \eqref{bond_length_distribution_single_gaussian_chain_approximated}
has a maximum at $Q / \sqrt{N_{0} b^{2}} = 1 / \sqrt{6}$.

As shown in the main text, the distribution functions for the single chain
non-interacting slip-link model are qualitatively different from ones
for the single chain repulsive slip-link model.
Such differences arise only from the difference of the interaction
between neighboring slip-links.

\section{Calculations of Integrals}
\label{detail_calculations_for_integrals}

In this appendix, we show the detailed calculations for several integrals
appear during the derivation of distribution functions in the main text.
Although the calculations are complicated, we can obtain the analytic
expressions by using special functions such as the gamma function or the
modified Bessel function.

\subsection{Integrals over $\lbrace N_{k} \rbrace$}
\label{integrals_over_n_k}

First we show the detailed calculation of the integral which appears in
eq \eqref{n_integral_modified}.
\begin{equation}
 \label{n_integral_modified_i1_definition}
 I_{1} \equiv \frac{1}{N^{5 Z / 2 - 1}} \int d\lbrace N_{k} \rbrace \,
   \delta \bigg( N - \sum_{k = 1}^{Z} N_{k} \bigg)
   \prod_{k = 1}^{Z} N_{k}^{3/2}
\end{equation}
To calculate the integral over $\lbrace N_{k} \rbrace$, we introduce
the following variable transform.
\begin{equation}
 s_{k} =
  \begin{cases}
   0 & (k = 0) \\
   (N_{k} - N_{k - 1}) / N & (k = 1,2,\dots,Z - 1) \\
   1 & (k = Z)
  \end{cases}
\end{equation}
The integral over $\lbrace
N_{k} \rbrace$ with the delta function can be transformed into
integral over $\lbrace s_{k} \rbrace$ without a delta function.
\begin{equation}
 \label{n_transform}
  \frac{1}{N^{Z - 1}} \prod_{k = 1}^{Z} \int_{0}^{N} dN_{k} \,
  \delta \bigg( N - \sum_{k = 1}^{Z} N_{k} \bigg)
  = \prod_{k = 1}^{Z - 1} \int_{0}^{s_{k + 1}} ds_{k}
\end{equation}

Eq \eqref{n_integral_modified_i1_definition} can be modified as follows,
by using the transform \eqref{n_transform}.
\begin{equation}
 \label{n_integral_modified_detail}
   I_{1}
   = 
   \prod_{k = 1}^{Z - 1} \int_{0}^{s_{k + 1}} ds_{k} \,
   \prod_{l = 0}^{Z - 1} (s_{l + 1} - s_{l})^{3/2}
\end{equation}
We consider the integral over $s_{1}$ in eq
\eqref{n_integral_modified_detail}. It can be modified as
\begin{equation}
 \int_{0}^{s_{2}} ds_{1} \, (s_{2} -
  s_{1})^{3/2} s_{1}^{3/2}
  = B(5/2,5/2) s_{2}^{4}
\end{equation}
where $B(x,y)$ is the beta function \cite{Abramowitz-Stegun-book}.
Similarly, the integral over $s_{2}$ in eq
\eqref{n_integral_modified_detail} can be calculated to be
\begin{equation}
 \int_{0}^{s_{3}} ds_{2} \, (s_{3} -
  s_{2})^{3/2} \left[  B(5/2,5/2) s_{2}^{4} \right]
  = B(5/2,5/2) B(5/2,10/2) s_{3}^{13/2}
\end{equation}
Iterating the same procedure, finally we have the following expression
for $I_{1}$.
\begin{equation}
 \label{n_integral_detail_modified}
   I_{1}
   = \prod_{k = 1}^{Z - 1} B(5/2,5 k / 2)
   = \prod_{k = 1}^{Z - 1} \frac{\Gamma(5/2) \Gamma(5 k /
   2)}{\Gamma(5 (k + 1) / 2)}
   = \frac{\Gamma^Z(5/2)}{\Gamma(5 Z / 2)}
\end{equation}
where we have utilized the relation between the beta and gamma
functions \cite{Abramowitz-Stegun-book}.
\begin{equation}
 B(x,y) = \frac{\Gamma(x) \Gamma(y)}{\Gamma(x + y)}
\end{equation}
By substituting $\Gamma(5/2) = 3 \sqrt{\pi} / 4 $ into eq
\eqref{n_integral_detail_modified}, we have eq \eqref{n_integral_modified}.

Next we show the detailed calculation of the integral in eq
\eqref{n_integral_with_constraint_modified}.
\begin{equation}
 \label{n_integral_with_constraint_modified_detail}
  \begin{split}
   I_{2} & \equiv \frac{1}{N^{5 Z / 2 - 2}} \int
   d\lbrace N_{k} \rbrace \, \left[ \frac{1}{Z} \sum_{l = 1}^{Z} \delta(n - N_{l})
   \right]
   \delta \bigg( N - \sum_{k = 1}^{Z} N_{k} \bigg)
  \prod_{k = 1}^{Z} N_{k}^{3/2} \\
   & = \frac{1}{Z} \sum_{l = 1}^{Z} \prod_{k = 1}^{Z - 1} \int_{0}^{s_{k
   + 1}} ds_{k} \,
   \delta(\tilde{n} - (s_{l + 1} - s_{l})) \prod_{m = 0}^{Z - 1} (s_{m + 1} - s_{m})^{3/2} \\
  \end{split}
\end{equation}
where we defined $\tilde{n} \equiv n / N$.
The integral over $\lbrace s_{k} \rbrace$ can be performed in
the similar way to the case of $I_{1}$.
\begin{equation}
 \label{n_integral_with_constraint_modified_detail_modified2}
  \begin{split}
   I_{2}
   & = \frac{1}{Z} \sum_{l = 1}^{Z} \prod_{k = l + 1}^{Z - 1} \int_{0}^{s_{k
   + 1}} ds_{k} \, \prod_{m = l + 1}^{Z - 1} (s_{m + 1} -
   s_{m})^{3/2} \\
   & \qquad \times \int_{0}^{s_{l + 1}} ds_{l} \,
   \delta(\tilde{n} - (s_{l + 1} - s_{l}))
   (s_{l + 1} - s_{l})^{3/2} s_{l}^{(5 l - 2) / 2} \prod_{m = 1}^{l - 1} B(5m/2,5/2)\\
   & = \frac{\tilde{n}^{3/2}}{Z} \sum_{l = 1}^{Z} \prod_{k = l + 1}^{Z - 1} \int_{\tilde{n}}^{s_{k
   + 1}} ds_{k} \, \prod_{m = l + 1}^{Z - 1} (s_{m + 1} -
   s_{m})^{3/2}
   (s_{l + 1} - \tilde{n})^{(5 l - 2) / 2} \prod_{m = 1}^{l - 1} B(5m/2,5/2)\\
  \end{split}
\end{equation}
To simplify the expression, we introduce another variable transform,
$u_{k} = s_{k} - \tilde{n}$. Then we have
\begin{equation}
 \label{n_integral_with_constraint_modified_detail_modified3}
  \begin{split}
   I_{2} 
   & = \frac{\tilde{n}^{3/2}}{Z} \sum_{l = 1}^{Z} \prod_{k = l + 1}^{Z - 1} \int_{0}^{\tilde{u}_{k
   + 1}} du_{k} \, \prod_{m = l + 1}^{Z - 1} (u_{m + 1} -
   u_{m})^{3/2}
   u_{l + 1}^{(5 l - 2) / 2} \prod_{m = 1}^{l - 1} B(5m/2,5/2) \\
   & = \frac{\tilde{n}^{3/2}}{Z} \sum_{l = 1}^{Z} (1 - \tilde{n})^{[5 (Z - 1) - 2] / 2} \prod_{m = 1}^{Z - 2} B(5m/2,5/2) \\
  \end{split}
\end{equation}
Finally we have the following expression for $I_{2}$.
\begin{equation}
 \label{n_integral_with_constraint_modified_detail_modified4}
   I_{2} 
   = \tilde{n}^{3/2} (1 - \tilde{n})^{5 (Z - 1) / 2 - 1}
   \frac{\Gamma^{Z - 1}(5 / 2)}{\Gamma(5 (Z - 1) / 2)}
\end{equation}
Eq \eqref{n_integral_with_constraint_modified_detail_modified4} together
with $\tilde{n} = n / N$ and $\Gamma(5 / 3) = 3 \sqrt{\pi} / 4$ gives eq \eqref{n_integral_with_constraint_modified}.

\subsection{Integrals over $n$}
\label{integrals_over_n}

Here we show the detailed calculation of eq
\eqref{bond_distribution_single_chain_approximated}. The integral over
$n$ in eq \eqref{bond_distribution_single_chain_approximated} can be
modified as follows, by using the variable transform $t = \sqrt{{5 b^{2}} / {3 N_{0}
 \bm{Q}^{2}}} n$.
\begin{equation}
 \label{bond_distribution_single_chain_approximated_detail}
  I_{3} \equiv \frac{1}{N_{0}}
  \int_{0}^{\infty} dn \,
  \exp \left( - \frac{3 \bm{Q}^{2}}{2 n b^{2}} - \frac{5}{2}
  \frac{n}{N_{0}} \right)
  = \sqrt{\frac{3 \bm{Q}^{2}}{5 N_{0} b^{2}}} \int_{0}^{\infty} dt \,
   \exp \bigg[ - \frac{1}{2} \sqrt{\frac{15 \bm{Q}^{2}}{N_{0} b^{2}}}
  \left(
  \frac{1}{t} +
  t \right) \bigg]
\end{equation}
The integral in eq \eqref{bond_distribution_single_chain_approximated_detail} can
be reduced to the first order modified Bessel function of the
second kind. The integral expression of the first order modified Bessel
function of
the second kind becomes\cite{Abramowitz-Stegun-book}
\begin{equation}
 \label{integral_expression_k1}
  K_{1}(x)
  = \int_{0}^{\infty} du \, e^{ - x \cosh u} \cosh u
  = \frac{1}{2} \int_{-\infty}^{\infty} du \, e^{u - x \cosh u}
  = \frac{1}{2} \int_{0}^{\infty} ds \,
  \exp \left[ - \frac{x}{2} \left(s + \frac{1}{s} \right) \right]
\end{equation}
where we have used the variable transform $s = e^{-u}$.
By substituting $x = \sqrt{15 \bm{Q}^{2} / N_{0} b^{2}}$ into eq
\eqref{integral_expression_k1},
we have the following expression for $I_{3}$.
\begin{equation}
 \label{bond_distribution_single_chain_approximated_detail_modified2}
  I_{3}
  = \frac{2}{5} \sqrt{\frac{15 \bm{Q}^{2}}{N_{0} b^{2}}} K_{1} \bigg(
  \sqrt{\frac{15 \bm{Q}^{2}}{N_{0} b^{2}}} \bigg)
\end{equation}
This gives eq \eqref{bond_distribution_single_chain_approximated}.

A similar integral also appears in eq
\eqref{bond_distribution_single_gaussian_chain_modified}.
The integral in eq
\eqref{bond_distribution_single_gaussian_chain_modified} can be
calculated as follows.
\begin{equation}
 \label{bond_distribution_single_gaussian_chain_modified_modified}
 \begin{split}
 I_{4}
  & \equiv
  \frac{N - N_{0}}{N_{0} N} \int_{0}^{N} dn \,
  \left(\frac{N_{0}}{n}\right)^{3/2}
  \exp \left[ - \frac{3 \bm{Q}^{2}}{2 n b^{2}} - \frac{N - N_{0}}{N_{0} N} n \right] \\
  & = \frac{2 (Z_{0} - 1)}{Z_{0}^{3/2}} \int_{1}^{\infty} dt \,
  \exp \left[ - \frac{3 \bm{Q}^{2}}{2 N b^{2}} t^{2} -
  ( Z_{0} -
   1 ) \frac{1}{t^{2}} \right]
 \end{split}
\end{equation}
where we used the variable transform $t = \sqrt{N / n}$.
The integral over $t$ in eq
\eqref{bond_distribution_single_gaussian_chain_modified_modified} can be
calculated by using the following formula \cite{Abramowitz-Stegun-book}.
\begin{equation}
  \int_{1}^{\infty} dt \, e^{-x t^{2} - y / t^{2}} \\
  = \frac{\sqrt{\pi}}{4 \sqrt{x}}
  \left[ e^{2 \sqrt{xy}} \erfc(\sqrt{x} + \sqrt{y})
  + e^{- 2 \sqrt{xy}} \erfc(\sqrt{x} - \sqrt{y}) \right]
\end{equation}
where $\erfc x$ is the complementary error function \cite{Abramowitz-Stegun-book}.
Finally eq \eqref{bond_distribution_single_gaussian_chain_modified_modified} can
be reduced to
\begin{equation}
 \label{bond_distribution_single_gaussian_chain_modified_modified2}
 \begin{split}
 I_{4}
  & = \frac{\sqrt{\pi} (Z_{0} - 1)}{Z_{0}^{3/2}}
  \sqrt{\frac{N b^{2}}{6 \bm{Q}^{2}}}
  \bigg[ \exp\bigg[\sqrt{\frac{6 \bm{Q}^{2}}{N b^{2}}
  ( Z_{0} -  1 )}\bigg] \erfc\bigg(\sqrt{\frac{3 \bm{Q}^{2}}{2 N b^{2}}} + \sqrt{Z_{0} -
   1}\bigg) \\
  & \qquad + \exp\bigg[- \sqrt{\frac{6 \bm{Q}^{2}}{N b^{2}}
  ( Z_{0} -  1 )}\bigg] \erfc\bigg(\sqrt{\frac{3 \bm{Q}^{2}}{2 N b^{2}}} - \sqrt{Z_{0} -
   1}\bigg) \bigg]
 \end{split} 
\end{equation}
This gives eq \eqref{bond_distribution_single_gaussian_chain_modified}.

\section{Calculations of Saddle Point Approximations}
\label{saddle_point_approximations_for_distribution_functions}

\subsection{Saddle Point Approximation for {\rm $P_{\text{eq}}(Z)$}}
\label{saddle_point_approximation_for_p_z}

A simple approximate expression is demanding to analyze the statistical
properties of the repulsive slip-link model.
Here we attempt to calculate the approximate
expression for eq
\eqref{sliplinked_distribution_single_chain_modified}, for sufficiently
large $Z_{0}$. For convenience, we approximate $P_{\text{eq}}(Z)$ by a
continuum distribution. The summation over $Z$ is replaced by the
integral as follows.
\begin{equation}
 \sum_{Z = 1}^{\infty} \approx \int_{-\infty}^{\infty} dZ
\end{equation}

We can utilize the Stirling's formula
\cite{Abramowitz-Stegun-book} to
approximate the gamma function.
\begin{equation}
  \frac{\xi^{Z - 1}}{\Gamma(5
   Z / 2)}
  \approx \exp \left[ (Z - 1) \ln \xi
  - \left(\frac{5 Z}{2} - \frac{1}{2}\right) \ln \frac{5 Z}{2} + \frac{5 Z}{2} - \frac{1}{2} \ln (2 \pi)
  \right]
\end{equation}
Next we expand the exponent around the saddle point.
\begin{equation}
 \label{sliplinked_distribution_single_chain_exponent_of_integrand}
 \begin{split}
  f(Z)
  & \equiv (Z - 1) \ln \xi
  - \left( \frac{5 Z}{2} - \frac{1}{2} \right) \ln \frac{5 Z}{2} + \frac{5 Z}{2} -
  \frac{1}{2} \ln (2 \pi) \\
  & \approx f(Z^{*}) + \frac{1}{2} f''(Z^{*}) (Z - Z^{*})^{2}
 \end{split}
\end{equation}
where $Z^{*}$ is the saddle point value of $Z$, which satisfies the
saddle point equation.
\begin{equation}
 \label{sliplinked_distribution_single_chain_saddle_point_equation}
  f'(Z^{*})
  = \ln \xi
  - \frac{5}{2} \ln \frac{5 Z^{*}}{2} + \frac{1}{2 Z^{*}}
  = 0
\end{equation}
The distribution function \eqref{sliplinked_distribution_single_chain_modified} can
be approximated as a Gaussian.
\begin{equation}
 \label{sliplinked_distribution_single_chain_saddle_point_approximation}
 \begin{split}
  P_{\text{eq}}(Z)
  & \approx \frac{\displaystyle \exp \left[
  f''(Z^{*}) (Z - Z^{*})^{2} / 2 \right]}{\displaystyle \int_{-\infty}^{\infty} dZ \, \exp \left[
  f''(Z^{*}) (Z - Z^{*})^{2} / 2 \right]} \\
  & = \sqrt{\frac{5 Z^{*} + 1}{4 \pi (Z^{*})^{2}}} \exp \left[
  - \frac{5 Z^{*} + 1}{4 (Z^{*})^{2}} (Z - Z^{*})^{2} \right]
 \end{split}
\end{equation}
The average value of $Z$ is calculated to be
\begin{equation}
 \label{sliplinked_distribution_single_chain_saddle_point_value}
 \langle Z \rangle_{\text{eq}} = Z_{0}
  \approx Z^{*}
\end{equation}
Therefore we can replace $Z^{*}$ in eq
\eqref{sliplinked_distribution_single_chain_saddle_point_approximation}
by $Z_{0}$. Finally we have eq
\eqref{sliplinked_distribution_single_chain_approximated} as the
approximate form for the subchain number distribution function.

\subsection{Saddle Point Approximation for {\rm $P_{\text{eq}}(n)$}}
\label{saddle_point_approximation_for_p_n}

We consider to obtain a simple approximate expression for eq
\eqref{segment_distribution_single_chain_definition_modified2}.
For sufficiently large $Z_{0}$, we can use the following
approximate form.
\begin{equation}
  \begin{split}
   \sum_{Z = 2}^{\infty}
   \left( 1 - \frac{n}{N} \right)^{5 (Z - 1) / 2}
   \frac{\xi^{Z - 1}}{\Gamma(5 (Z - 1) / 2)}
   & \approx \int_{-\infty}^{\infty} dZ \,
   \exp \left[
   g(Z^{**}) +
   \frac{1}{2} g''(Z^{**}) (Z - Z^{**})^{2}\right] \\
   & = \sqrt{\frac{2 \pi}{g''(Z^{**})}} e^{g(Z^{**})} \\
 \end{split}
\end{equation}
where we defined
\begin{equation}
  g(Z) \equiv
  \frac{5 Z}{2} \ln \left( 1 - \frac{n}{N} \right) + Z \ln \xi
  - \frac{5 Z - 1}{2} \ln \frac{5 Z}{2} + \frac{5 Z}{2} -
  \frac{1}{2} \ln (2 \pi)
\end{equation}
and $Z^{**}$ is given via the following saddle point equation.
\begin{equation}
 g'(Z^{**}) = \frac{5}{2} \ln \left( 1 - \frac{n}{N} \right) + \ln \xi
  - \frac{5}{2} \ln \frac{5 Z^{**}}{2} + \frac{1}{2 Z^{**}} = 0
\end{equation}
From eqs \eqref{sliplinked_distribution_single_chain_saddle_point_equation}
and \eqref{sliplinked_distribution_single_chain_saddle_point_value},
we have the following approximate relation between
$Z^{**}$ and $Z_{0}$.
\begin{equation}
 Z^{**} \approx \left( 1 - \frac{n}{N} \right) Z_{0}
\end{equation}
Then we can write
\begin{equation}
   \sum_{Z = 2}^{\infty}
   \left( 1 - \frac{n}{N} \right)^{5 (Z - 1) / 2}
   \frac{\tilde{\xi}^{Z - 1}}{\Gamma(5 (Z - 1) / 2)} \\
   \approx \frac{Z_{0}}{\sqrt{e}} \left( 1 - \frac{n}{N} \right)
   \exp \left[ \frac{5 Z_{0}}{2} \left( 1 - \frac{n}{N} \right) \right] \\
\end{equation}
and the segment number distribution function can be expressed approximately as
\begin{equation}
 \label{segment_distribution_single_chain_definition_approximated_unnormalized}
 \begin{split}
  P_{\text{eq}}(n)
   & \propto
  \left( \frac{n}{N} \right)^{3/2}
  \left( 1 - \frac{n}{N} \right)^{-1}
  \left[ \left( 1 - \frac{n}{N} \right)
   \exp \left[ \frac{5 Z_{0}}{2} \left( 1 - \frac{n}{N}
  \right) \right] \right] \\
   & \propto n^{3/2}
   \exp \left( - \frac{5 n}{2 N_{0}} \right) \\
 \end{split}
\end{equation}
By normalizing eq
\eqref{segment_distribution_single_chain_definition_approximated_unnormalized},
finally we have eq
\eqref{segment_distribution_single_chain_definition_approximated_final}
as the approximate form for $P_{\text{eq}}(n)$.

\section{Strong Repulsion Limit}
\label{equilibrium_distribution_functions_at_strong_repulsion_limit}

In the main text and Appendix
\ref{equilibrium_distribution_functions_for_entangled_signle_gaussian_chain},
we calculated some equilibrium distribution functions for
repulsive and non-interacting slip-link models, respectively.
For comparison, here we consider the case where the repulsive
interaction between slip-links is quite strong. We call such a limit
as the strong repulsion limit. As we will show, slip-links form a Wigner
crystal like ordered structure in this limit.

We assume the following effective potential for slip-links, instead of
eq \eqref{effective_potential_pcn_sliplink}.
\begin{equation}
 \label{effective_potential_pcn_sliplink_strong_repulsion_limit}
  \tilde{U}_{\text{slip-link}}(n) = - \alpha k_{B} T \ln n
\end{equation}
Here, $\alpha$ is the parameter which represents to the strength of the
repulsive interaction. ($\alpha = 0$ and $\alpha = 3/2$ corresponds to
the non-interacting and repulsive slip-link models, respectively.)
At the strong repulsion limit, we set $\alpha \to \infty$.

The equilibrium probability distribution function and the grand
partition function are expressed as follows.
\begin{equation}
 \begin{split}
  P_{\text{eq}}(\lbrace \bm{R}_{k} \rbrace,\lbrace N_{k} \rbrace,Z)
   & = \frac{e^{\epsilon Z / k_{B} T}}{\Xi \Lambda^{3 (Z + 1)} N^{Z -
  1}} \delta\bigg(N - \sum_{k = 1}^{Z} N_{k}\bigg)
  \prod_{k = 1}^{Z} N_{k}^{\alpha - 3 / 2}
  \exp\left[ - \frac{3 (\bm{R}_{k} - \bm{R}_{k - 1})^{2}}{2 N_{k} b^{2}}
  \right]  
 \end{split}
\end{equation}
\begin{equation}
 \begin{split}
 \Xi
  & \equiv \sum_{Z = 1}^{\infty} \frac{e^{\epsilon Z / k_{B} T}}{\Lambda^{3 (Z + 1)} N^{Z - 1}}
  \int d\lbrace \bm{R}_{k} \rbrace d\lbrace N_{k} \rbrace \,
  \delta\bigg(N - \sum_{k = 1}^{Z} N_{k}\bigg)
\prod_{k = 1}^{Z} N_{k}^{\alpha - 3 / 2}
  \exp\left[ - \frac{3 (\bm{R}_{k} - \bm{R}_{k - 1})^{2}}{2 N_{k} b^{2}}
  \right]
 \end{split}
\end{equation}
Since the repulsive interaction energy becomes very large at the strong
repulsion limit, we can reasonably
utilize the saddle point approximation for $\lbrace N_{k} \rbrace$.
For $\alpha \gg 1$, the distribution of $\lbrace N_{k} \rbrace$ is expected to be
sufficiently sharp and thus
the fluctuations around the saddle point is negligible.
Then the grand
partition function can be approximated as
\begin{equation}
 \begin{split}
 \Xi
  & \approx \sum_{Z = 1}^{\infty} \frac{e^{\epsilon Z / k_{B} T}}{\Lambda^{3 (Z + 1)} N^{Z - 1}}
  \int d\lbrace \bm{R}_{k} \rbrace \,
  \left(\frac{N}{Z}\right)^{(\alpha - 3 / 2) Z}
  \exp\left[ - \frac{3 Z (\bm{R}_{k} - \bm{R}_{k - 1})^{2}}{2 N b^{2}}
  \right] \\
  & = \frac{\mathcal{V} N}{\Lambda^{3}} \sum_{Z = 1}^{\infty}
  \bigg[
  \left(\frac{2 \pi b^{2}}{3 \Lambda^{2}}\right)^{3/2}
  \frac{N^{\alpha - 1}}{Z^{\alpha}} e^{\epsilon / k_{B} T}
  \bigg]^{Z}
 \end{split}
\end{equation}

For simplicity, we assume that the average number of subchains is
sufficiently large ($Z_{0} \gg 1$) and
utilize the saddle point approximation for $Z$. The result is
\begin{equation}
 \label{sliplinked_distribution_strong_repulsion_limit_approximated}
 \begin{split}
  P_{\text{eq}}(Z)
  & \approx \left(\frac{\alpha}{2 \pi Z_{0}}\right)^{1/2} \exp\left[ - \frac{\alpha}{2 Z_{0}} (Z - Z_{0})^{2}
  \right] \\
  & \to \delta(Z - Z_{0}) \qquad (\alpha \to \infty)
 \end{split}
\end{equation}
Thus we find that the subchain number distribution simply
becomes the delta function. This means that there is essentially no
subchain number fluctuation in the strong repulsion limit.
Similarly, the segment number distribution function is obtained as
\begin{equation}
 \label{segment_number_distribution_strong_repulsion_limit}
 P_{\text{eq}}(n) \approx \delta(n - N_{0})
\end{equation}
Eq \eqref{segment_number_distribution_strong_repulsion_limit} means that 
the chemical distance between two neighboring slip-links is constant
(slip-links are placed equidistantly on a chain).
Eqs
\eqref{sliplinked_distribution_strong_repulsion_limit_approximated}
and \eqref{segment_number_distribution_strong_repulsion_limit} are
naturally derived from the dependence of the variances of $Z$ and $n$ on
$\alpha$. From the results in Appendix
\ref{equilibrium_distribution_functions_for_entangled_signle_gaussian_chain},
the main text, and this appendix, the variances are roughly estimated as $\langle (Z - Z_{0})^{2} \rangle_{\text{eq}} \approx Z_{0} / (1 +
\alpha)$ and $\langle (n - N_{0})^{2} \rangle_{\text{eq}} \approx N_{0}^{2} /
(1 + \alpha)$. At the strong repulsion limit ($\alpha \to 0$), both of
them approach to zero and the distribution functions reduce to delta
functions (as eqs
\eqref{sliplinked_distribution_strong_repulsion_limit_approximated} and \eqref{segment_number_distribution_strong_repulsion_limit}).
We can interpret such a state as a sort of Wigner
crystal. This situation seems to be qualitatively different from most
of slip-links models such as the PCN model.
It would be rather similar to the simple tube
model\cite{Doi-Edwards-book} in which each tube segment contains a
constant number of segments. (But of course, the slip-link model is not equivalent to the tube model.)
Anyway, the distribution functions for $Z$ and $n$ at the strong
repulsion limit are much sharper than
distribution functions for the non-interacting or repulsive slip-link models.
Thus we consider that these distribution functions become sharper as the
repulsive interaction between slip-links increases.

Because the segment number $n$ is fixed to $N_{0}$, the bond vector
distribution function is nothing but a Gaussian.
The bond vector and bond length distribution functions become
\begin{align}
 & \label{bond_vector_distribution_strong_repulsion_limit}
 P_{\text{eq}}(\bm{Q}) \approx \left(\frac{3}{2 \pi N_{0} b^{2}}\right)^{3/2}
 \exp\left(-\frac{3 \bm{Q}^{2}}{2 N_{0} b^{2}} \right) \\
 & \label{bond_length_distribution_strong_repulsion_limit}
 P_{\text{eq}}(Q)
 \approx \frac{4}{\sqrt{\pi}}
 \left(\frac{3}{2 N_{0} b^{2}}\right)^{3/2}  Q^{2} \exp\left(-\frac{3 Q^{2}}{2 N_{0} b^{2}} \right)
\end{align}
At the limit of small $Q$, we have $P_{\text{eq}}(Q) \propto Q^{2}$ as
the asymptotic form. This asymptotic
behavior is similar to one of the repulsive slip-link model (eq
\eqref{bond_length_distribution_single_chain_asymptotic}).
Eq \eqref{bond_length_distribution_strong_repulsion_limit} has a maximum
at $Q / \sqrt{N_{0} b^{2}}= \sqrt{2 / 3} \approx 0.816$, which is larger
than the value for the repulsive slip-link model. Thus
we expect that the value of $Q / \sqrt{N_{0} b^{2}}$ which gives the
maximum increases as the repulsive
interaction increases.


\clearpage

\section*{Figure Captions}

Figure \ref{sliplinked_number_distribution_function_single_chain_graph}:
The slip-linked subchain number distribution functions with various values of
$Z_{0}$. Solid and dashed curves show the exact and
approximate distribution functions for the single chain repulsive slip-link model,
respectively.
The dotted curves show the Poisson distribution which
corresponds to the distribution of the single chain non-interacting
slip-link model.
$Z_{0} = 5, 10, 20, 40,$ and $80$ from left to right.
The approximate distributions almost coincide to the exact distributions.

\vspace{\baselineskip}
\hspace{-\parindent}%
Figure \ref{average_and_variance_of_sliplinked_number_graph}:
Dependence of the variance of slip-linked subchain number $\langle (Z -
Z_{0})^{2} \rangle_{\text{eq}}$ on the average $\langle Z \rangle_{\text{eq}} = Z_{0}$. The solid
and dashed curves show the exact and approximate variances for
the repulsive slip-link model, and the dotted curve shows the variance for
the non-interacting slip-link model ($\langle (Z -
Z_{0})^{2} \rangle_{\text{eq}} = Z_{0} - 1$).
The approximate curve is very close to the exact curve.

\vspace{\baselineskip}
\hspace{-\parindent}%
Figure \ref{segment_number_distribution_function_single_chain_graph}:
The segment number distribution functions for sufficiently large
$Z_{0}$. Solid and dotted curves show the distribution
functions for the repulsive and
non-interacting slip-link models, respectively.

\vspace{\baselineskip}
\hspace{-\parindent}%
Figure \ref{bond_length_distribution_function_single_chain_graph}:
The bond length distribution functions for sufficiently large
$Z_{0}$. Solid and dotted curves show the distribution
functions for the repulsive and
non-interacting slip-link models, respectively.

\vspace{\baselineskip}
\hspace{-\parindent}%
Figure \ref{sliplinked_number_distribution_function_comparison_graph}:
The slip-linked subchain number distribution functions by single chain
models and PCN simulations. Curves are theoretical predictions by single
chain models and symbols are simulation data.
$Z_{0} = 5, 10, 20,$ and $40$ from left to right.

\vspace{\baselineskip}
\hspace{-\parindent}%
Figure \ref{segment_number_distribution_function_comparison_graph}:
The segment number distribution functions calculated by single chain
models and PCN simulations (for $Z_{0} = 40$).
Curves and symbols represent the results by single chain
models and PCN simulations, respectively.

\vspace{\baselineskip}
\hspace{-\parindent}%
Figure \ref{bond_length_distribution_function_comparison_graph}:
The bond length distribution functions calculated by single chain
models and PCN simulations (for $Z_{0} = 40$).
Curves and symbols represent the results by single chain
models and PCN simulations, respectively.

\clearpage

\section*{Figures}

\vspace{7\baselineskip}


\end{document}